\documentclass{aa} 
\usepackage{psfig,graphics}

\begin{document}

   \thesaurus{03(11.08.1; 11.09.4; 11.19.2)} 

   \title{Spectroscopy of diffuse ionized gas in halos of selected edge-on 
          galaxies\thanks{Based on observations obtained at ESO/La Silla 
(Chile)}}
\titlerunning{Spectroscopy of extraplanar diffuse ionized gas}
   \author{R. T\"ullmann
     \and
           R.-J. Dettmar
}

   \offprints{tullmann@astro.ruhr-uni-bochum.de}
   \institute{Astronomisches Institut, Ruhr-Universit\"at Bochum,
              D-44780 Bochum, Germany}

   \date{Received 21 June 1999; accepted 14 July 2000}

   \maketitle

   \begin{abstract}
In order to examine the excitation and ionization mechanism of extraplanar 
diffuse ionized gas (DIG) we have obtained optical longslit spectra of seven 
edge-on spiral galaxies. In four objects the brightest emission lines can be 
traced out to distances of typically 1.5\,kpc above the disk.
For NGC\,1963 and NGC\,3044 line
ratios such as [\ion{N}{ii}]\,$\lambda6583/{\rm H\alpha}$ or 
[\ion{S}{ii}]\,$\lambda6717/{\rm H\alpha}$ as well as  
${\rm [\ion{O}{iii}]\,\lambda5007}$/${\rm H\beta}$ could be measured for 
the halo DIG. This allows us
to discuss the DIG in the halo of these objects in the framework
of diagnostic diagrams. For these two objects, the
line ratios of ${\rm [\ion{O}{iii}]\,\lambda5007}$/${\rm H\beta}$ 
decrease with increasing $|z|$, different from the recently 
reported trend in  NGC\,891 (Rand \cite{rand}).
We find that emission lines from the DIG in the disks are in good
agreement with photoionization models using a dilute radiation field. 
However, with 
increasing $|z|$ these models fail to predict the measured 
${\rm [\ion{O}{i}]\,\lambda 6300}$/${\rm H\alpha}$\ and
${\rm \ion{He}{i}\,\lambda 5876}$/${\rm H\alpha}$ line ratios for NGC\,1963.
Diagnostic diagrams reveal for NGC\,1963 the need for 
a second ionization mechanism of the halo DIG (besides photoionization).
 This additional source could be shock ionization. The same diagrams 
demonstrate an intermediate classification  for NGC\,3044. 
Plots of $[\ion{S}{ii}]/[\ion{N}{ii}]$ vs. emission measure reveal 
significant changes towards the halo and seem to trace local small scale 
density fluctuations of the extraplanar DIG.
   \end{abstract}

%
\section{Introduction}
Over the last decade a widespread and extended diffuse component 
of ionized hydrogen contributing 25\,--\,50\,\% to the total H$\alpha$ flux
 has been found well outside of classical 
\ion{H}{ii} regions in several face-on and edge-on galaxies (e.g., 
Lehnert\,\&\,Heckman \cite{lehe}, Rand \cite{rand96}, Greenawalt 
et al. \cite{green97},
or Dettmar \cite{det98}).
Although there is significant evidence that the presence 
of the extraplanar diffuse ionized gas (DIG, sometimes also called warm 
ionized medium, WIM) is correlated with ongoing star formation in the disk 
(e.g., Dettmar  \cite{det92}), important aspects of its 
ionization and excitation are not well understood. 
While of all conventional ionizing sources only
UV radiation of massive stars provides sufficient power
to keep these thick gas layers ionized (including the Reynolds-layer of 
the Milky Way (Reynolds \cite{rey93})) the resulting
photoionization models have to explain the long
mean free paths for the UV photons and the observed
emission line ratios. However, the relatively strong
emission observed in the  
$\rm{[\ion{N}{ii}]\,\lambda\lambda 6548,6584}$ lines 
(Dettmar\,\&\,Schulz \cite{deschu92}, Golla et al. \cite{gogode96}, Rand 
\cite{rand97}) cause a problem for pure photoionization models
and an additional heating contribution, e.g.  from the kinetic energy provided
by supernova, is likely (Dahlem et al. \cite{dahl97}).
An even more puzzling observation was recently presented
for the DIG in NGC\,891 which is up to now the best studied galaxy in 
this regard 
(Rand \cite{rand}). In complete contradiction to photoionization
models here [\ion{O}{iii}]/H$\beta$ is rising with distance above the galactic
plane for  $|z| >$\ 1\,kpc. Diagnostic diagrams for emission lines
from DIG in outflows of dwarf galaxies (Martin \cite{martin}) also demonstrate
the need for ionization and/or excitation processes beyond pure 
photoionization. 
In this paper we report results from  spectroscopic observations 
of a small sample of edge-on galaxies. These new spectra cover 
a larger wavelengths range extending earlier work to the blue range of the
optical spectrum. This allows us to discuss the results in the framework 
of diagnostic diagrams.
The paper is structured as follows: In Sect.~2--\,3
some essential information concerning observations and data reduction 
strategies, the determination of line ratios,  and the use of diagnostic 
diagrams is given. Sect.~4 shows representative results which are 
discussed briefly for each galaxy with respect to the photoionization models 
by Mathis (\cite{ma}, hereafter Ma86) and Domg\"orgen\,\&\,Mathis 
(\cite{doma}, hereafter DM94). Finally, our 
findings are summarized in Sect.~5 with special emphasis to the trend of 
$\rm{[\ion{S}{ii}]/[\ion{N}{ii}]}$, and the proposed dependence of line ratio 
variations with changing halo metallicities.
   
\section{Observations and data reduction}
The dataset consists of spectra of the normal late-type spirals 
NGC\,1963, IC\,2531, NGC\,3044, NGC\,4302, NGC\,4402, NGC\,4634, and of the 
irregular galaxy NGC\,2188. All objects in the sample are seen nearly edge-on 
(see Table~\ref{table1}).
Slit positions were determined from narrow band $\rm{H\alpha}$ images  on 
which diffuse \ion{H}{ii} gas at high galactic latitudes was detected 
(Rossa\,\&\,Dettmar \cite{rossa}).

The spectroscopic data were obtained during January/February 1995 with the 
ESO\,1.52m telescope in combination with the Boller\,\&\,Chivens spectrograph 
at La Silla (Chile). A Ford Aero\-space CCD (2048\,$\times$\,2048 pix\-el) 
with a pixel size of 15\,$\times$\,15\,$\mu$m and a spatial reso\-lution of 
0.82$\arcsec$\,pix$^{-1}$ was used during the observations. Gra\-ting 
$\#$\,23 provided a dispersion of 126\,\AA\,${\rm mm^{-1}}$ which resulted in 
a coverage of a wavelength range from 3500\,\AA\,--\,7450\,\AA. 
We determined the spectral resolution to 4.6\,\AA\ by measuring the FWHM of 
an unbroadened emission line in a HeAr-calibration spectrum. The total 
integration time per slit position was 90 minutes.

\begin{table*}
\caption[]{Sample overview}
\label{table1}
\begin{flushleft}
\begin{tabular}{c c c c c c c c c}
\hline \\
Galaxy & R.A.  & Dec. & Type & D  & z$^e$ & inclination  & $m_{R}$ & DIG 
morphology$^g$  \\ 
 & (J 2000) & (J 2000) & & [Mpc]$^a$ & & i$^f$ & [mag]$^d$ & \\
\\
\hline
\\ 
NGC\,1963  & 05$^{h}$33$^{m}$12.8$^{s}$ & $-$36$^{\circ}$23$'$59$''$ & 
Sc$^{\ \,}$  & 17.7$^{\ \ }$ & 0.004410 & $85^{\circ}$\,$^{\ }$ & 
12.11$^{\ }$ & diffuse \\

NGC\,3044 & 09$^{h}$53$^{m}$39.8$^{s}$ & +01$^{\circ}$34$'$46$''$ & 
SBb$^{\ \,}$ & 17.2$^{\ \ }$ & 0.004310 & $84^{\circ}$\,$^{b}$ & 
11.65$^{\ }$ & bright, diffuse\\
 
IC\,2531\hspace{0.35cm} & 09$^{h}$59$^{m}$55.7$^{s}$ & 
$-$29$^{\circ}$36$'$55$''$ & Sb$^{\ }$ & 33.0$^{\ \ }$ & 0.008460 & 
$90^{\circ}$\,$^{\ }$ & 11.54$^{\ }$ & 1 filament \\ 

NGC\,4302 & 12$^{h}$21$^{m}$42.4$^{s}$ & +14$^{\circ}$36$'$05$''$ & 
Sc$^{\ \,}$ & 18.8$^c$$^{\ }$ & 0.003690 & $88^{\circ}$\,$^{c}$ & 
11.83$^{c}$ & faint, diffuse\\ 

NGC\,4402 & 12$^{h}$26$^{m}$07.9$^{s}$ & +13$^{\circ}$06$'$46$''$ & Sb$^{\ }$ 
& 22.0$^{c}$$^{\ }$ & 0.000790 & $74^{\circ}$\,$^{c}$ & 12.09$^{c}$ & 
diffuse \\ 
NGC\,4634 & 12$^{h}$42$^{m}$40.4$^{s}$ & +14$^{\circ}$17$'$47$''$ & 
Sc$^{\ \, }$  & 19.1$^c$$^{\ }$ & 0.000390 & $83^{\circ}$\,$^{c}$ & 
11.91$^{c}$ & bright, diffuse, extraplanar \\ 
& & & & & & & & + filaments\\
\\
\hline
\end{tabular}
{\flushleft
{\footnotesize $^a \ {\rm for}\ H_{0} = 75\,{\rm km\,s^{-1} Mpc^{-1}}$},
{\footnotesize $^b$ Bottinelli et al. \cite{botti}},
{\footnotesize $^c$ Teerikorpi et al. \cite{teeri}}\\
{\footnotesize $^d$ The Surface Photometry Catalogue of the ESO-Uppsala 
Galaxies; Lauberts A. and Valentijn E.A., \cite{laube}}\\
{\footnotesize $^e$ NASA Extragalactic Database}\\
{\footnotesize $^f$ Third Reference Catalogue of Bright Galaxies; de 
Vaucouleurs et al., \cite{devau}}\\
{\footnotesize $^g$ Rossa\,\&\,Dettmar 2000}
}
\end{flushleft}
\end{table*}
 
Data reduction was performed using the IRAF and MIDAS software packages. All 
data were bias subtracted and dome flats were used to reduce small scale 
sensitivity variations. An extinction correction was performed by adopting 
the atmospheric extinction curve for La Silla (T\"ug \cite{tug}). 
Spectroscopy of the standard star Hiltner\,600 yielded the wavelength 
dependent response function. We divided all object frames by this function to 
correct for wavelength dependent sensitivity variations. Subsequently a flux 
calibration of our spectra was applied.

The slit length of $4\farcm2$ allowed a precise night sky subtraction. 
This reduction process did not influence the emission lines of 
\ion{He}{i}$\lambda$5876 and [\ion{O}{i}]$\lambda$6300 since they are 
sufficiently redshifted from the atmospheric lines for our objects.
Due to the large wavelength range we could not apply a single continuum fit 
to the whole spectrum. 
Therefore the continuum level was estimated and subtracted row by row
from the total emission of each galaxy using linear fits to the continuum 
on each side of prominent emission lines. 
The data for NGC\,2188 are already presented by Dom\-g\"or\-gen\,\&\,Dett\-mar 
(\cite{domade}) and will not be further discussed here. As a consistency
check these spectra were reanalysed and the previous results were fully 
reproduced.

\section{Line ratios and diagnostic diagrams}
For all galaxies listed in Table~1  emission lines are traced down to
S/N\,$=$\,$2\sigma$.
Detected spectral features for all galaxies are 
$\rm{[\ion{N}{ii}]\,\lambda\lambda
6548,6583}$, $\rm{H\alpha}$ and $\rm{[\ion{S}{ii}]\,\lambda\lambda6717,6732}$.
Line ratios are measured using a block average algorithm.
The block size was fixed by the achieved S/N 
(at least 2) to reach simultaneously the best possible spatial resolution.
For the disk and the halo of a galaxy we have chosen block sizes of 
1\,--\,2 and 3\,--\,4 rows, respectively.

For the galaxies NGC\,1963 and NGC\,3044 additional emission lines could be
detected from the halo DIG, namely 
$\rm{[\ion{O}{ii}]\,\lambda 3727}$\footnote{Due to low spectral resolution 
the [\ion{O}{ii}]-doublet cannot be 
separated. We therefore assume an average wavelength of 3727\,\AA.}, 
$\rm{H\beta}$, $\rm{[\ion{O}{iii}]\,\lambda\lambda 4959,5007}$, 
$\rm{\ion{He}{i}\,\lambda 5876}$, and $\rm{[\ion{O}{i}]\,\lambda 6300}$.
Thus the extraplanar DIG in these objects can be discussed for the
first time in the framework of diagnostic diagrams. A typical spectrum
of the DIG in NGC\,1963 is given in Fig.~2.

A comparison with photoionization models of Ma86 and DM94 can clarify some 
properties of the sources which keep the DIG ionized.
Only the line ratios of 
$\rm{[\ion{O}{ii}]\,\lambda 3727}$/$\rm{[\ion{O}{iii}]\,\lambda 5007}$ 
and $\rm{[\ion{O}{iii}]\,\lambda 5007}/{\rm H\beta}$ have to be 
corrected for interstellar extinction. The precise values 
are obtained following Osterbrock (\cite{oster}). Unreddened line ratios
can be computed by assuming an average interstellar extinction curve and 
adopting $\rm{H\beta}$ as nebular reference emission line. The extinction can
 be determined from the measured Balmer decrement. 
It is assumed that such a correction is negligible for the remaining line 
ratios because only pairs of emission lines in the red spectral region had 
been chosen which are located in a narrow wavelength region ($< 200$\,\AA) 
so that differential reddening is not significant.

\addtocounter{figure}{1}
\begin{figure*}[!pht]
\psfig{file=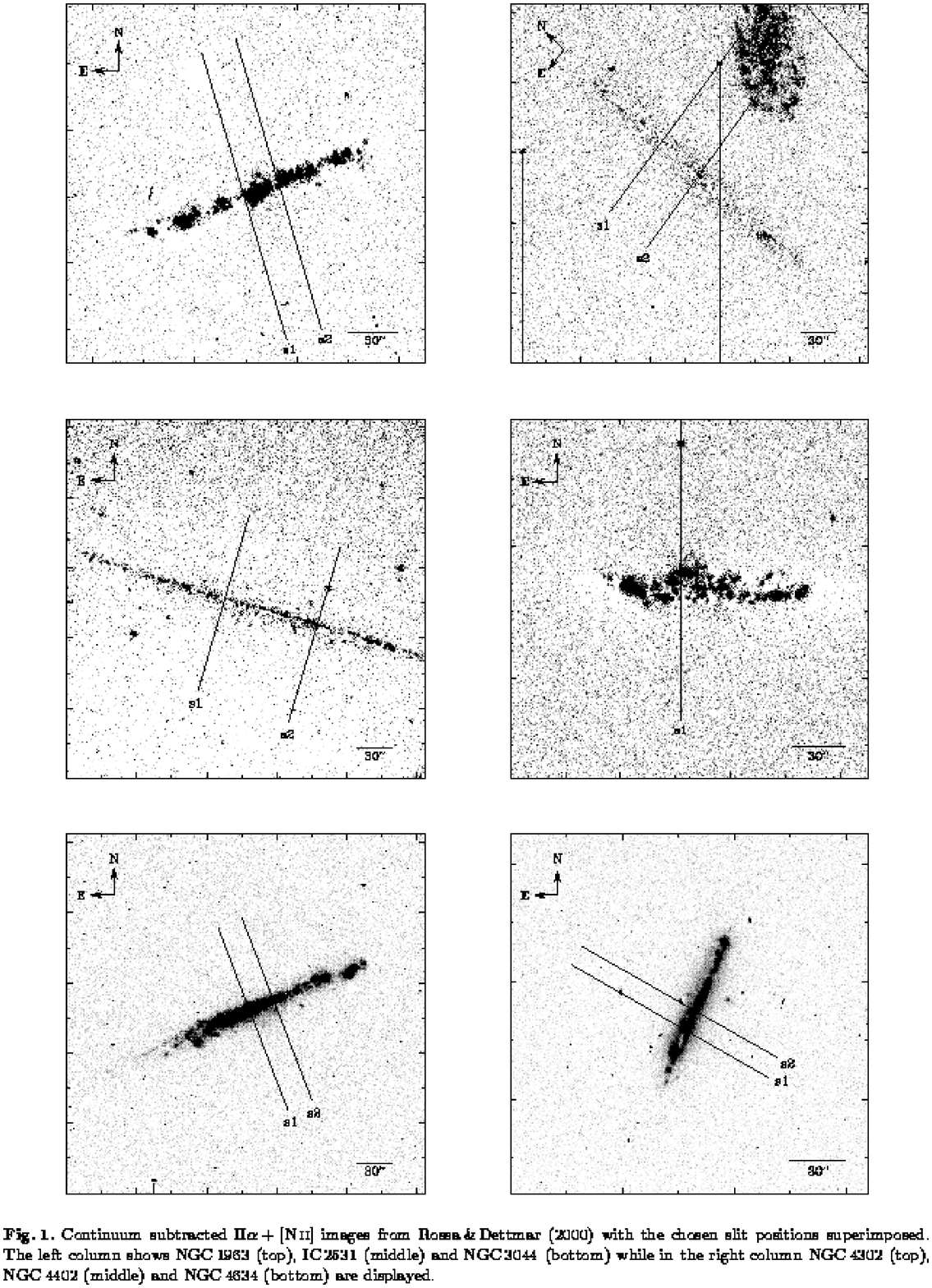,height=23cm,width=18cm}
\label{fig1}
\end{figure*}



%

It is noteworthy that an extinction correction does not influence a clear 
separation of different excitation mechanisms significantly 
(Baldwin et al. \cite{bald}).
The dereddening for 
$\rm{[\ion{O}{ii}]\,\lambda 3727}$/$\rm{[\ion{O}{iii}]\,\lambda 5007}$ 
and $\rm{[\ion{O}{iii}]\,\lambda 5007}/{\rm H\beta}$ amounts to 11$\%$ for 
the disk emission and 4$\%$ for the halo emission. For the remaining line
ratios the reddening correction is less than 1$\%$ (disk and halo).

In extragalactic objects basically four different ionization mechanisms can 
be clearly distinguished. Firstly ``normal'' photoionization by O stars 
which is present mainly in \ion{H}{ii} regions (star forming regions). 
Secondly shock ionization present in objects which are excited by supernova 
remnants (SNR) or by superwinds. 
Finally we can also distinguish between photoionization by 
extremely hot stars such as the central stars of planetary nebulae and 
objects which are 
photoionized by a power-law continuum source (AGN). To probe the ionization 
mechansism(s) of the extraplanar DIG it is useful to plot line ratios by 
pairs on a 
logarithmic scale (Baldwin et al. \cite{bald}, Veilleux\,\&\,Osterbrock 
\cite{veio}). 
Since the above mentioned mechanisms produce characteristically different 
spectra, all objects cover different locations when plotted into a single 
diagnostic diagram.

In order to verify our findings obtained from line ratios and diagnostic 
diagrams we examined also the broadening of line widths at extraplanar
latitudes. Therefore we plotted a kinematical diagnostic diagram (line ratio 
vs. line width (FWHM)) which indicates the presence of additional excitation 
mechanisms, such as shocks or stellar winds.

\section{Results and discussion}
\subsection{NGC\,1963}
 NGC\,1963 is a Sc galaxy with an inclination angle of $85^{\circ}$ (edge-on) 
at a distance of 17.7~Mpc (see Table~\ref{table1}). The corresponding 
slit positions s1 and s2 are overplotted on the ${\rm H\alpha}$ image  and are 
displayed in Fig.~\ref{fig1}. 
Some diffuse emission near the nucleus in the disk-halo interface and bright 
extended \ion{H}{ii} regions in the disk are visible. 
The origin of the $z$-scale was determined from the continuum emission of the 
disk. 

\begin{figure}
\hspace{-0.2cm}
\psfig{file=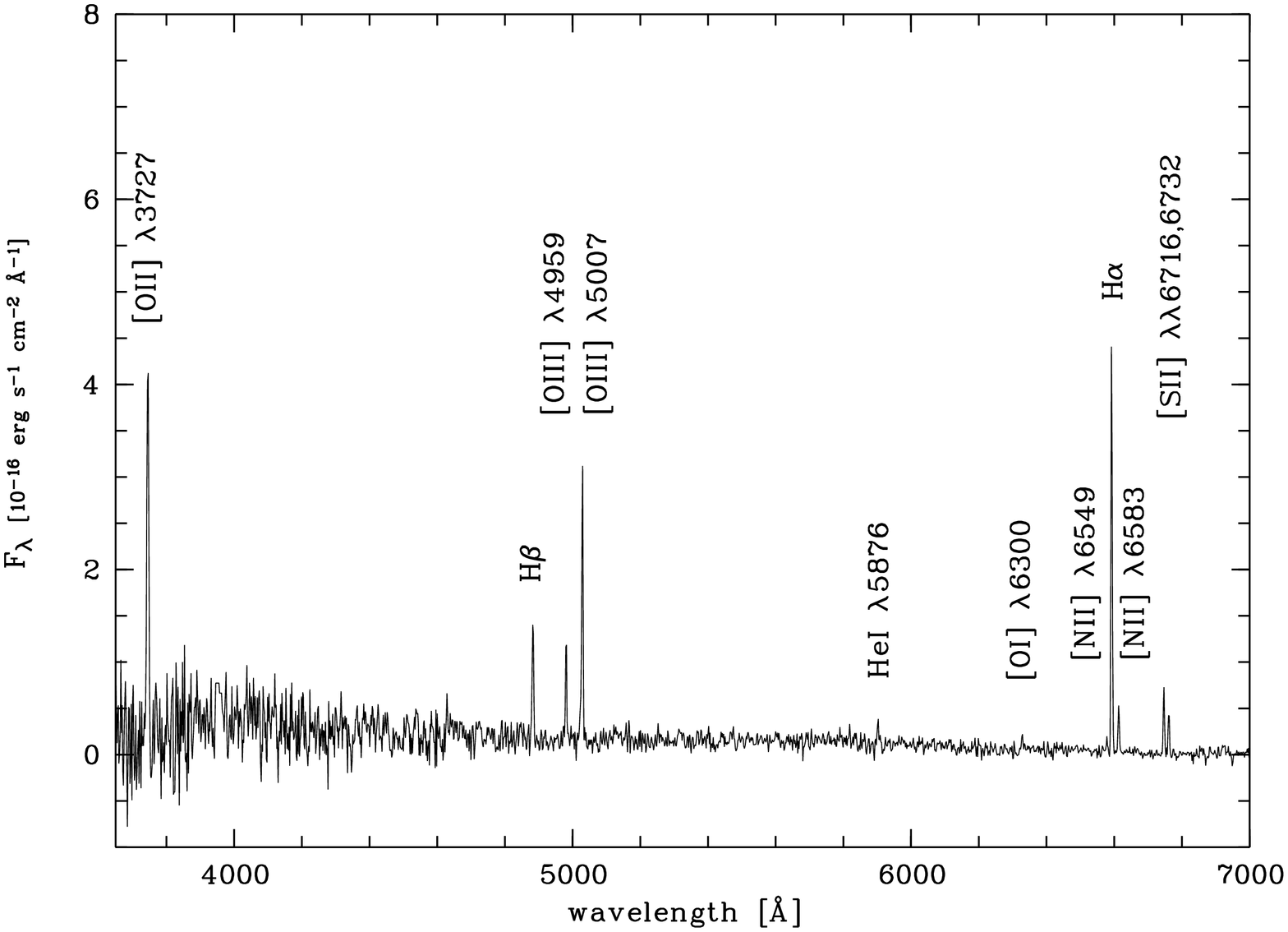,height=6.5cm,width=8.8cm,angle=-90}
\caption{NGC\,1963: Representative intensity plot (z\,=\,3$\arcsec$, 
corresponding to 260\,pc)
 for slit position 
s1. Rest wavelengths (in [\AA]) of relevant emission lines are marked. This 
spectrum shows 
the line and continuum emission.}
\label{fig2}
\end{figure}

\begin{figure*}[!ph]
\begin{minipage}[t]{18cm}
\setlength{\unitlength}{1cm}
\begin{picture}(10,8.8)(0,1)
\hspace{0.25cm}
\psfig{file=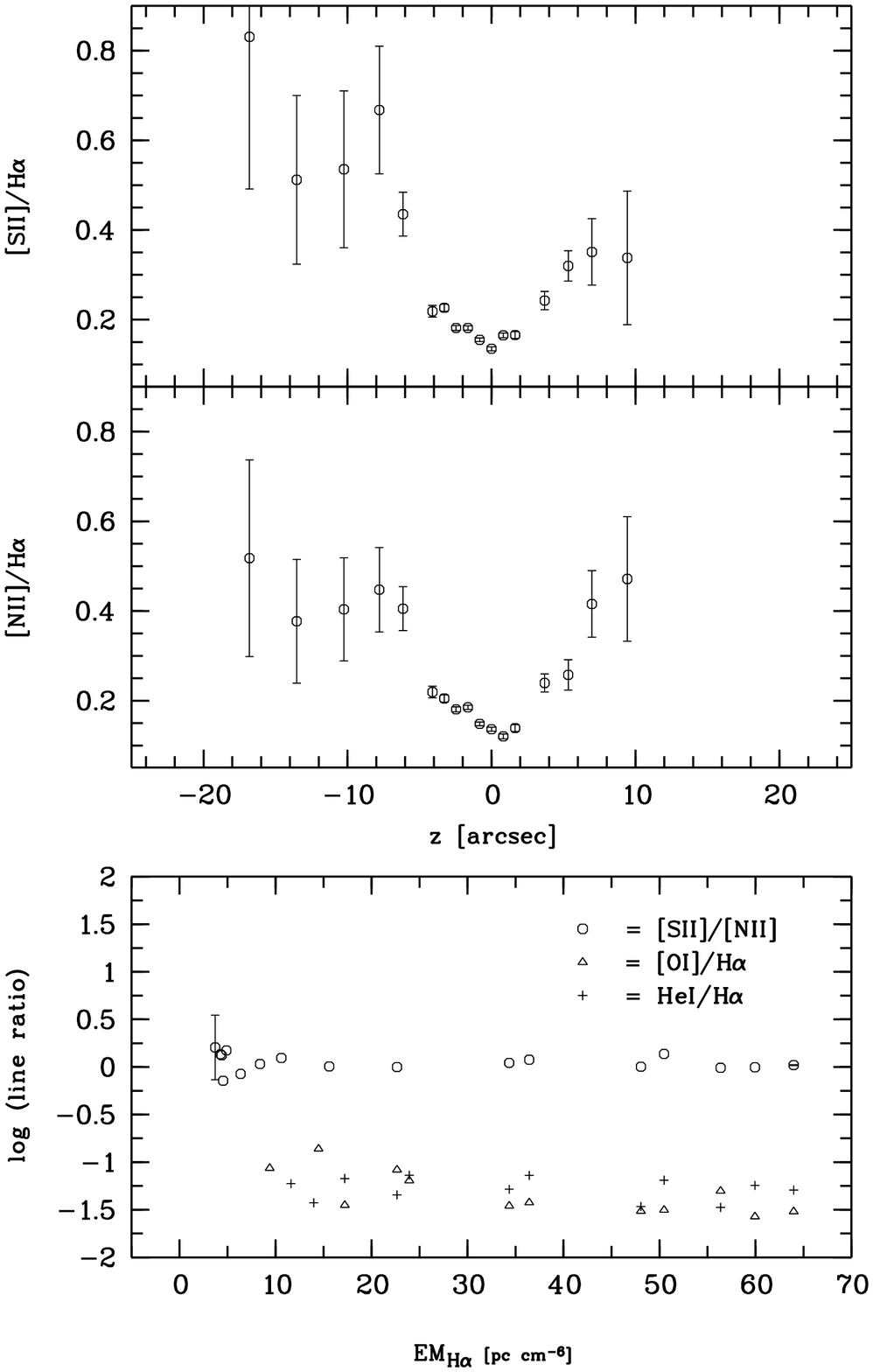,height=9cm,width=8cm,clip=t,angle=0}
\hspace{1cm}
\psfig{file=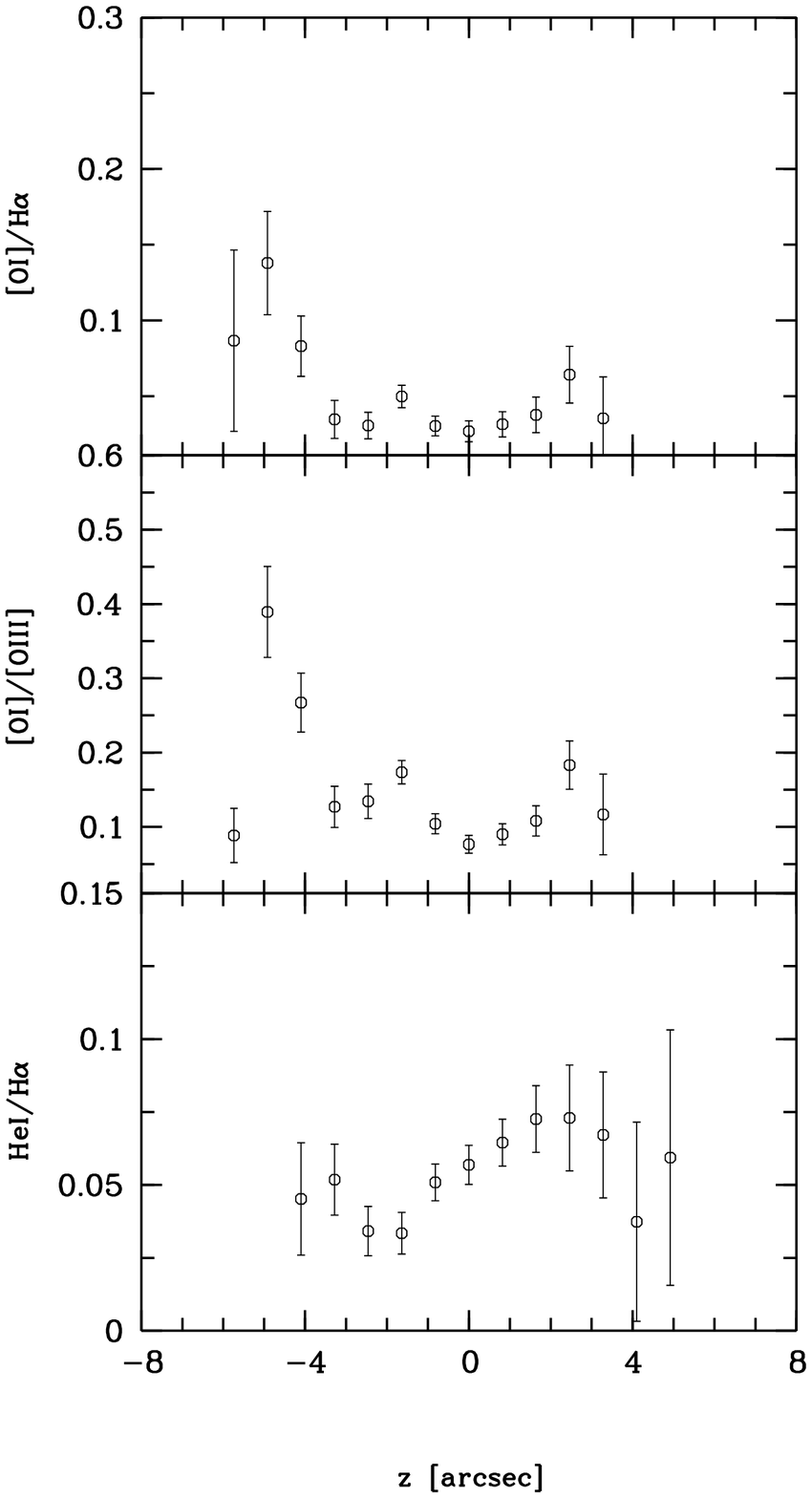,height=9cm,width=8cm,clip=t,angle=0}
\end{picture}
\end{minipage}
\vspace{0.5cm}
\\

\parbox[b]{18cm}{\caption{NGC\,1963: Line ratios of 
[\ion{S}{ii}]\,$\lambda$6717\,/\,H$\alpha$, 
[\ion{N}{ii}]\,$\lambda$6583\,/\,H$\alpha$ 
(left) and [\ion{O}{i}]\,$\lambda$6300\,/\,H$\alpha$, 
[\ion{O}{i}]\,$\lambda$6300\,/\,[\ion{O}{iii}]\,$\lambda$5007, 
and \ion{He}{i}\,$\lambda$5876\,/\,H$\alpha$ (right) along slit position s1. 
5$\arcsec$ correspond to 430\,pc.
The lower left panel displays the variation of line ratios as a function of
 emission measure EM$_{\rm H\alpha}$. Representative mean errors of 
[\ion{S}{ii}]/[\ion{N}{ii}] (halo and disk) are overplotted.}
\label{fig3}}

\begin{minipage}[t]{18cm}
\setlength{\unitlength}{1cm}
\begin{picture}(10,8.8)(0,1.75)
\hspace{0.25cm}
\psfig{file=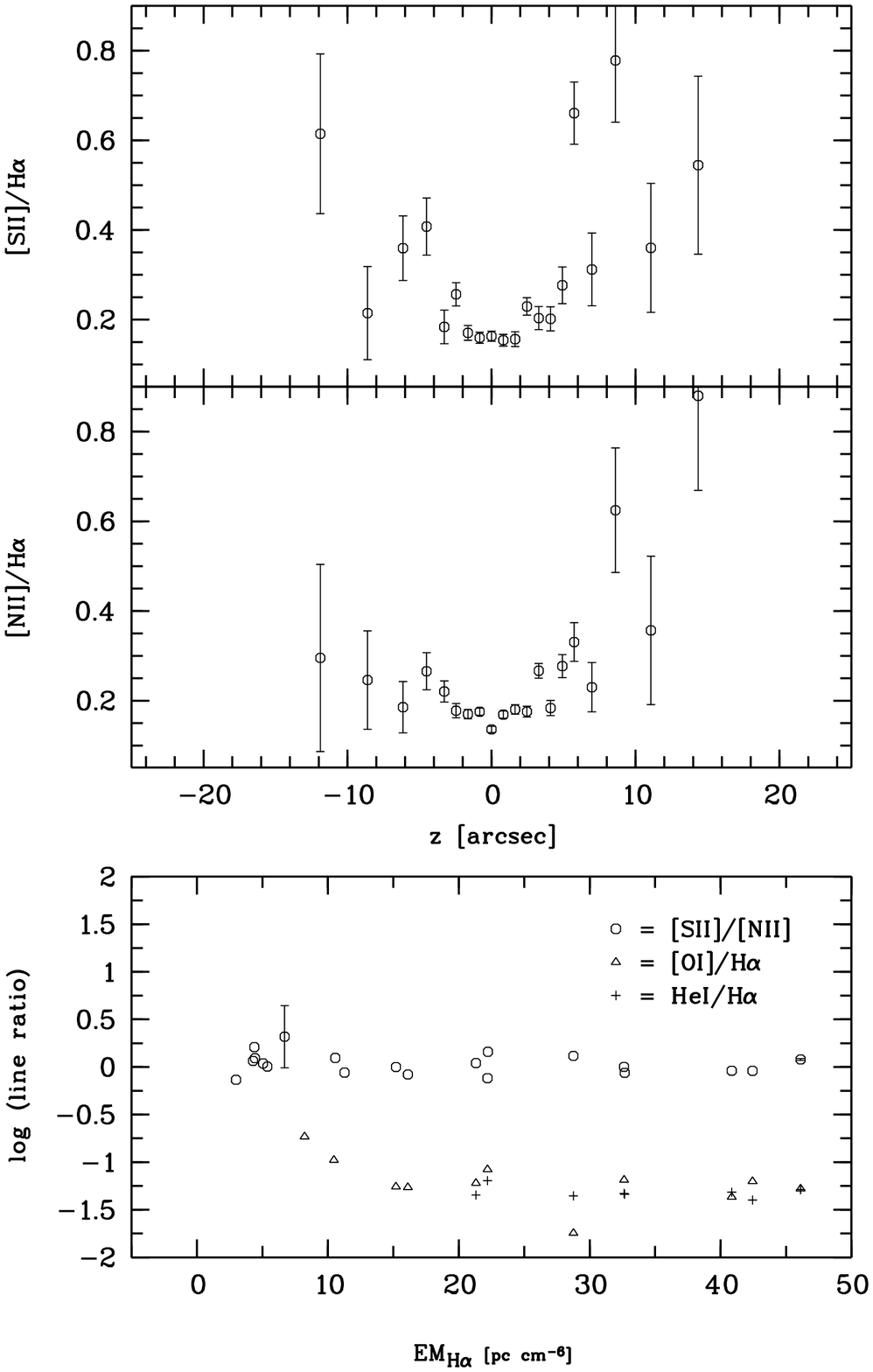,height=9.4cm,width=8cm,clip=t,angle=0}
\hspace{1cm}
\psfig{file=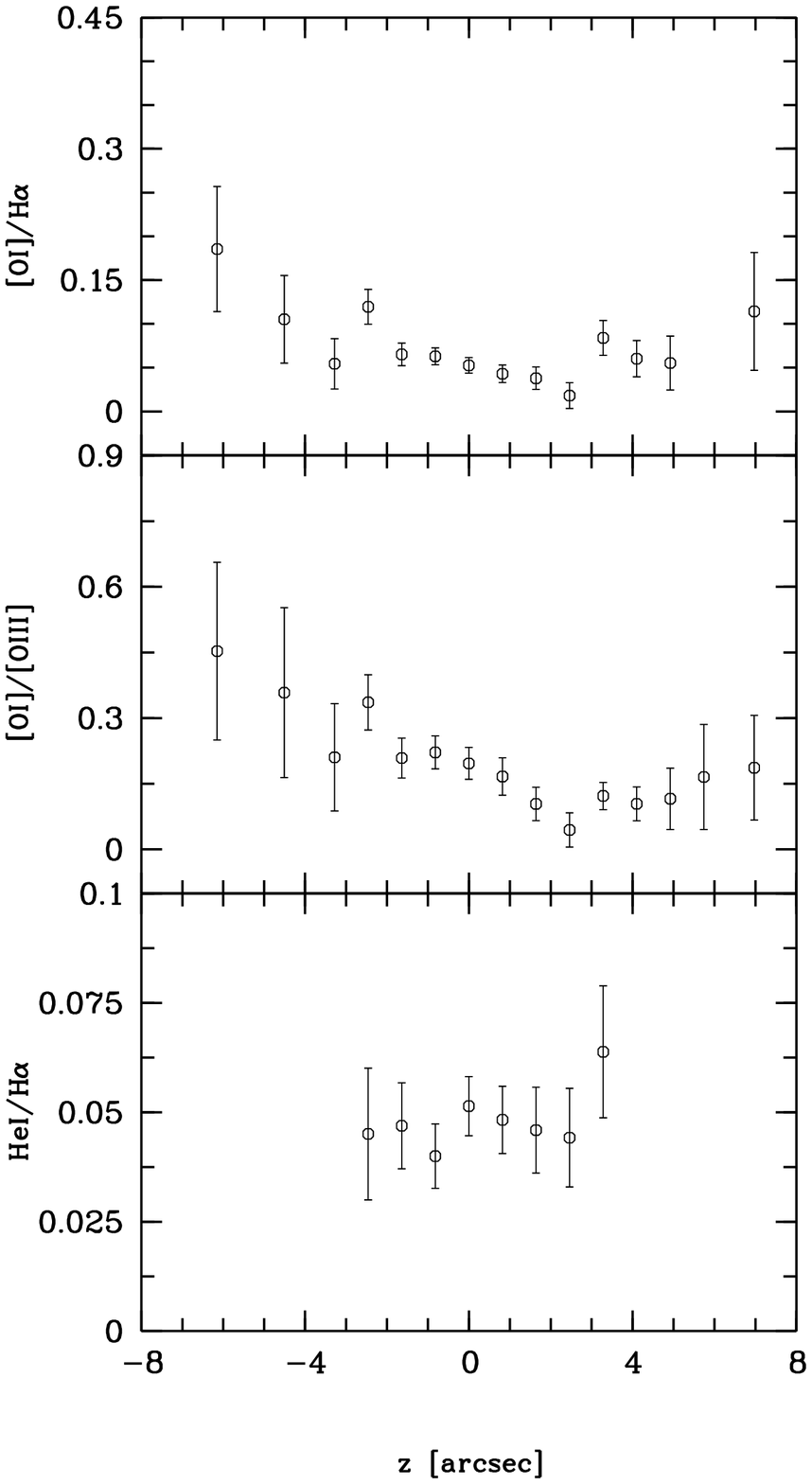,height=9.4cm,width=8cm,clip=t,angle=0}
\end{picture}
\end{minipage}
\vspace{1cm}
\\

\parbox[b]{18cm}{\caption{NGC\,1963: Same ratios of 
[\ion{S}{ii}]\,$\lambda$6717\,/\,H$\alpha$, 
[\ion{N}{ii}]\,$\lambda$6583\,/\,H$\alpha$ 
(left) and [\ion{O}{i}]\,$\lambda$6300\,/\,H$\alpha$, 
[\ion{O}{i}]\,$\lambda$6300\,/\,[\ion{O}{iii}]\,$\lambda$5007,
 and \ion{He}{i}\,$\lambda$5876\,/\,H$\alpha$ (right) but this time for slit 
position s2.  Plots in the lower left are 
the same as for s1. Representative mean errors (halo/disk) are plotted for 
[\ion{S}{ii}]/[\ion{N}{ii}]. Note the tight correlation between 
[\ion{O}{i}]/H$\alpha$ and \ion{He}{i}/H$\alpha$ in the lower left panels
of each figure.}
\label{fig4}}
\end{figure*}

\begin{figure*}[!ph]
\begin{minipage}[t]{18cm}
\setlength{\unitlength}{1cm}
\begin{picture}(10,8.8)(0,-2.7)
\hspace{0.25cm}
\psfig{file=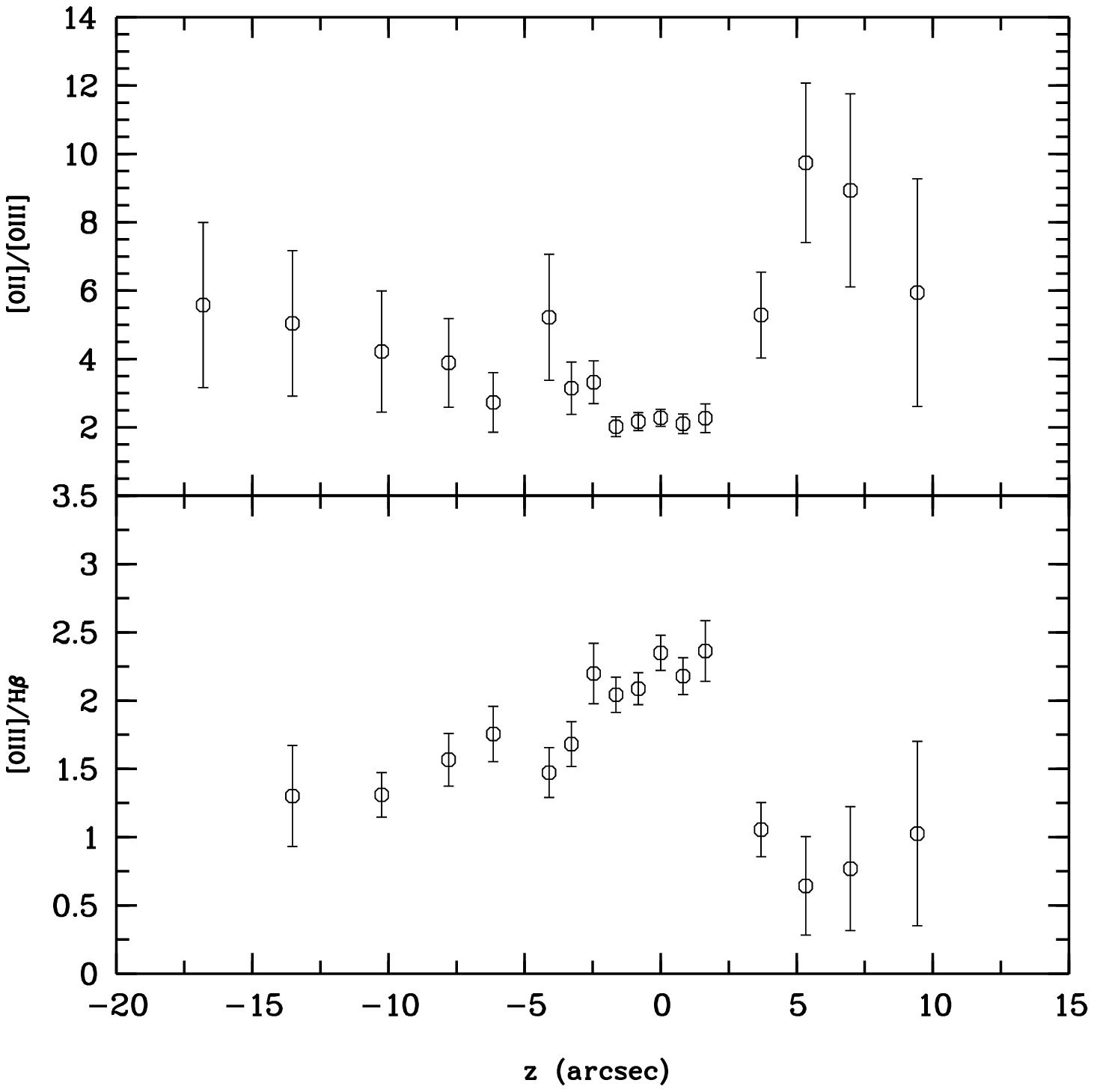,height=6cm,width=8cm,clip=t,angle=0}
\hspace{1cm}
\psfig{file=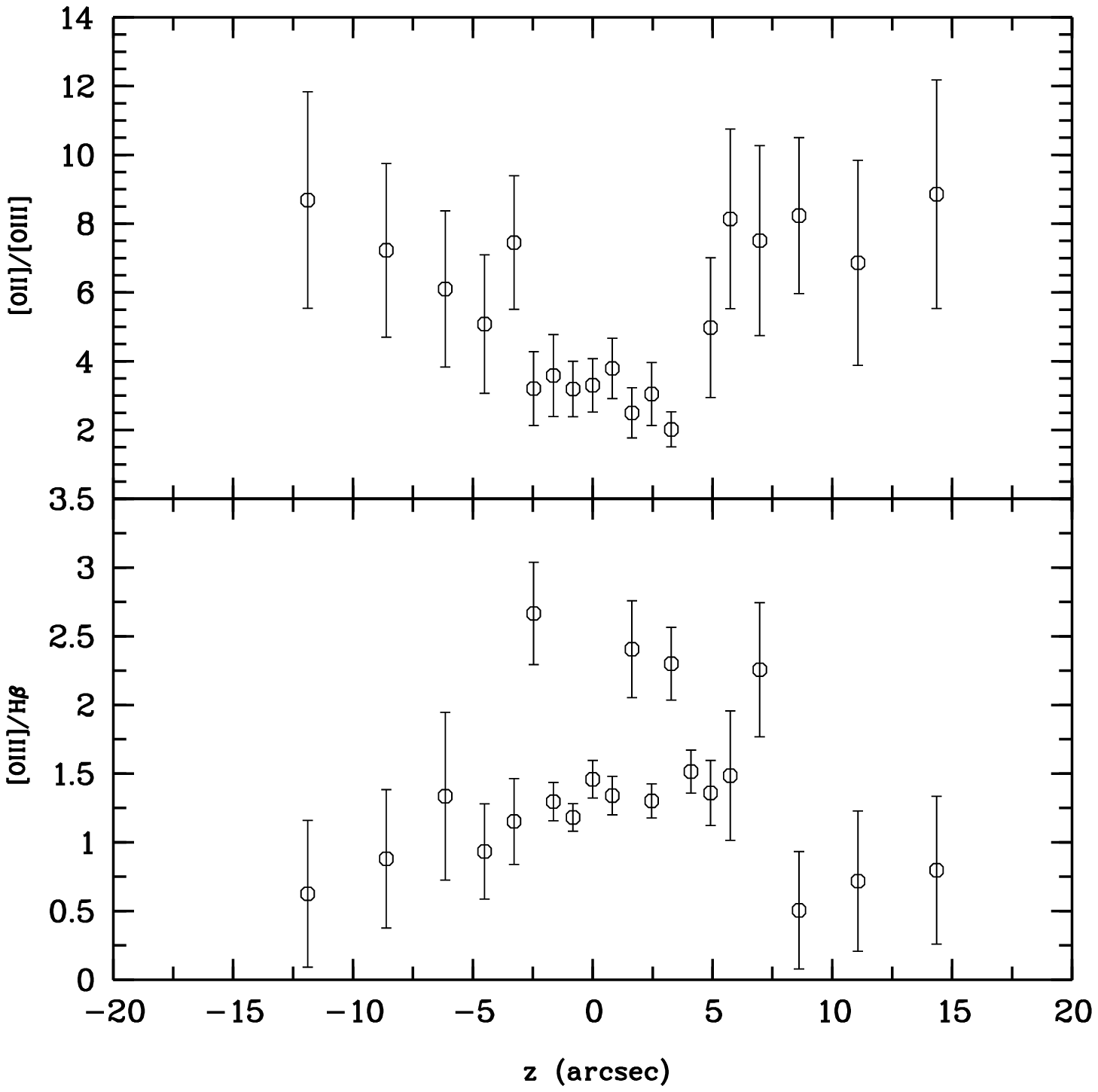,height=6cm,width=8cm,clip=t,angle=0}
\end{picture}
\end{minipage}
\vspace{-2.7cm}

\parbox[b]{18cm}{\caption{NGC\,1963: Measured line ratios of 
[\ion{O}{ii}]\,$\lambda$3727\,/\,[\ion{O}{iii}]\,$\lambda$5007 and 
[\ion{O}{iii}]\,$\lambda$5007\,/\,H$\beta$ along slit position s1 (left) and 
s2 
(right) .}
\label{fig5}}
\begin{center}
\hspace{0.0cm}
\psfig{file=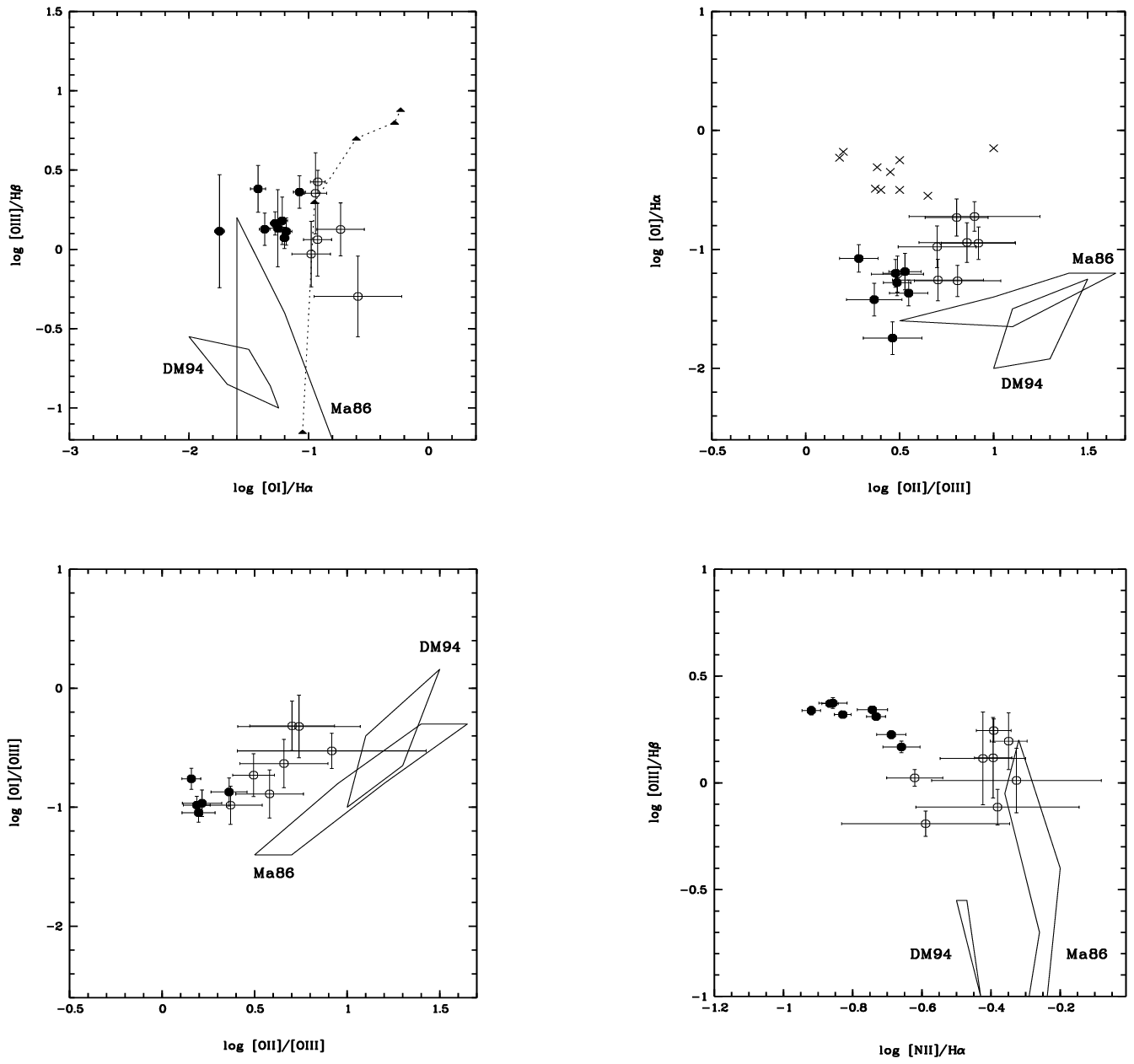,height=13cm,width=15cm,angle=0}
\end{center}
\vspace{-0.4cm}
\caption{NGC\,1963: 
Relevant diagnostic diagrams for slit position s1. The dotted line and
filled triangles represent the shock model from Shull\,\&\,McKee 
(\cite{shull}) with shock velocities between 80\,--\,100\,${\rm km\ s^{-1}}$.
Filled circles denote the disk data and open circles address the halo 
component. Crosses represent observed shock ionized objects used by Baldwin 
et al. (\cite{bald}). Areas enclosed by solid lines are fitted by Ma86 and 
DM92.
}
\label{fig6}
\end{figure*}

Although generally plotted as a function of $z$, we also present line ratios 
as a function of emission measure (EM). This provides additional information 
about the dependence of line ratios and the mean squared densities of free 
electrons in the emitting region.
Line ratios for positions s1 and s2 are plotted in 
Figs.~\ref{fig3}\,--\,\ref{fig5}. Table~\ref{table2} gives an overview of 
averaged line ratios for the disk, the northern, and the southern halo.
Representative logarithmic diagnostic diagrams for slit position s1 are 
displayed in Fig.~\ref{fig6}. 
Diffuse $\rm{H\alpha}$ emission can be traced spectroscopically in the 
northern halo out to 1.4\,kpc and out to 1.3\,kpc in the south. 
The line ratios [\ion{N}{ii}]/$\rm{H\alpha}$ and [\ion{S}{ii}]/$\rm{H\alpha}$ 
measured near the midplane of NGC\,1963 are typical for \ion{H}{ii} regions 
and show no conspicuous features. Both line ratios increase towards the outer 
halo ($\rm{|z|> 500\,pc}$). They are well fitted by the DM94 model taking into account a composite model with 
20\%\,($X_{edge} = 0.95$)\,+\,80\%\,($X_{edge} = 0.1)$.
$X_{edge}$ is the fraction of neutral hydrogen at the outer border of 
the model nebula. If $X_{edge}$ is high (radiation bounded case) nearly all 
emitted energy is absorbed by the geometry and only few stellar photons 
ionize hydrogen at the edge. A lower value of $X_{edge}$ (matter bounded case)
 indicates that a significant fraction of ionizing photons escapes the 
geometry.	

At both slit positions the averaged value for \ion{He}{i}/$\rm{H\alpha}$ is 
$0.05\pm0.01$ (cf. Table~\ref{table2}). With respect to the error bars the 
DM94 model fails in predicting the correct ratios no 
matter which parameter set is chosen. This is also true for the 
shock sensitive line ratio [\ion{O}{i}]/$\rm{H\alpha}$ at position s2.
It follows that there is most likely a second excitation mechanism besides 
photoionization. As the diagnostic diagrams with [\ion{O}{i}]$\lambda$6300 
emission (see Fig.~\ref{fig6}) reveal, this extraplanar source 
could be shock ionization.

\begin{table}[t]
\caption{Averaged line ratios for the disk and halo of NGC\,1963. Halo 
values 
are given for distances perpendicular to the plane as indicated below.}

\label{table2}
\begin{minipage}[t]{8.8cm}
\begin{tabular}{l l l l}
\hline
\vspace{-0.25cm}
\\

slit 1 & disk & halo (north) & halo (south) \\
& & $z=1.3$\,kpc & $z=0.7$\,kpc \\
\vspace{-0.25cm}
\\
\hline
\vspace{-0.2cm}
& & & \\

[\ion{N}{ii}]/${\rm H\alpha}$ & 0.16 $\pm$ 0.01 & 0.46 $\pm$ 0.25 & 0.36 
$\pm$ 0.09\\ 

[\ion{S}{ii}]/${\rm H\alpha}$ & 0.19 $\pm$ 0.01 & 0.63 $\pm$ 0.19 & 0.29 
$\pm$ 0.09 \\
 
\ion{He}{i}/${\rm H\alpha}$ & 0.05 $\pm$ 0.01 & \mbox{\hspace{0.55cm}} --- & 
\mbox{\hspace{0.55cm}} ---  \\

[\ion{O}{i}]/[\ion{O}{iii}] & 0.17 $\pm$ 0.03 & \mbox{\hspace{0.55cm}} --- & 
\mbox{\hspace{0.55cm}} --- \\ 

[\ion{O}{i}]/${\rm H\alpha}$ & 0.05 $\pm$ 0.01 & \mbox{\hspace{0.55cm}} --- & 
\mbox{\hspace{0.55cm}} ---\\ 

[\ion{O}{iii}]/${\rm H\beta}$ & 2.20 $\pm$ 0.18 & 1.53 $\pm$ 0.29 & 0.83 
$\pm$ 0.52\\ 

[\ion{O}{ii}]/[\ion{O}{iii}] & 2.06 $\pm$ 0.34 & 3.89 $\pm$ 1.57 & 6.57 $\pm$ 
2.00 \\ 
\vspace{-0.2cm}
& & & \\

\hline
\vspace{-0.25cm}
\\

 slit 2 &  disk  &  halo (north) &  halo (south) \\ 
& & $z=0.9$\,kpc & $z=1.0$\,kpc \\
\vspace{-0.25cm}
\\
\hline
\vspace{-0.2cm}
& & & \\

[\ion{N}{ii}]/${\rm H\alpha}$ & 0.18 $\pm$ 0.01 & 0.26 $\pm$ 0.12 & 0.53 
$\pm$ 0.19\\
 
[\ion{S}{ii}]/${\rm H\alpha}$ & 0.16 $\pm$ 0.01 & 0.44 $\pm$ 0.15 & 0.42 
$\pm$ 0.26\\

\ion{He}{i}/${\rm H\alpha}$ & 0.05 $\pm$ 0.01 & \mbox{\hspace{0.55cm}} --- & 
\mbox{\hspace{0.55cm}} --- \\

[\ion{O}{i}]/[\ion{O}{iii}] & 0.13 $\pm$ 0.04 & 0.33 $\pm$ 0.16 & 0.15 $\pm$ 
0.08 \\ 

[\ion{O}{i}]/${\rm H\alpha}$ & 0.04 $\pm$ 0.01 & 0.12 $\pm$ 0.05 & 0.09 $\pm$ 
0.05 \\ 

[\ion{O}{iii}]/${\rm H\beta}$ & 1.79 $\pm$ 0.23 & 0.78 $\pm$ 0.44 & 0.65 
$\pm$ 0.48\\

[\ion{O}{ii}]/[\ion{O}{iii}] & 2.90 $\pm$ 0.64 & 6.88 $\pm$ 2.51 & 6.91 $\pm$ 
2.63
\vspace{0.2cm} \\

\hline
\end{tabular}
\end{minipage}
\end{table}

The line ratios of [\ion{O}{iii}]/$\rm{H\beta}$ decrease with
increasing $|z|$ (Fig.~\ref{fig5}). This pattern 
is easily explained in photoinization models (e.g., Ma86) 
by the ionization stratification
expected from the ionization potentials involved.
However, recent observations in NGC\,891 (Rand \cite{rand}) show an increase 
of the line ratio of [\ion{O}{iii}]/$\rm{H\beta}$ with increasing distance
from the midplane and therefore from the most likely location of the
ionizing sources.

\subsection{IC\,2531}
This galaxy of type Sb is perfectly edge-on and shows no prominent DIG 
emission at either of the slit positions. 
Only a single filament can be detected 20\arcsec\ NE of slit s2 (cf. 
Rossa\,\&\,Dettmar \cite{rossa}). 

\begin{figure}[t]
\hspace{0.25cm}
\psfig{file=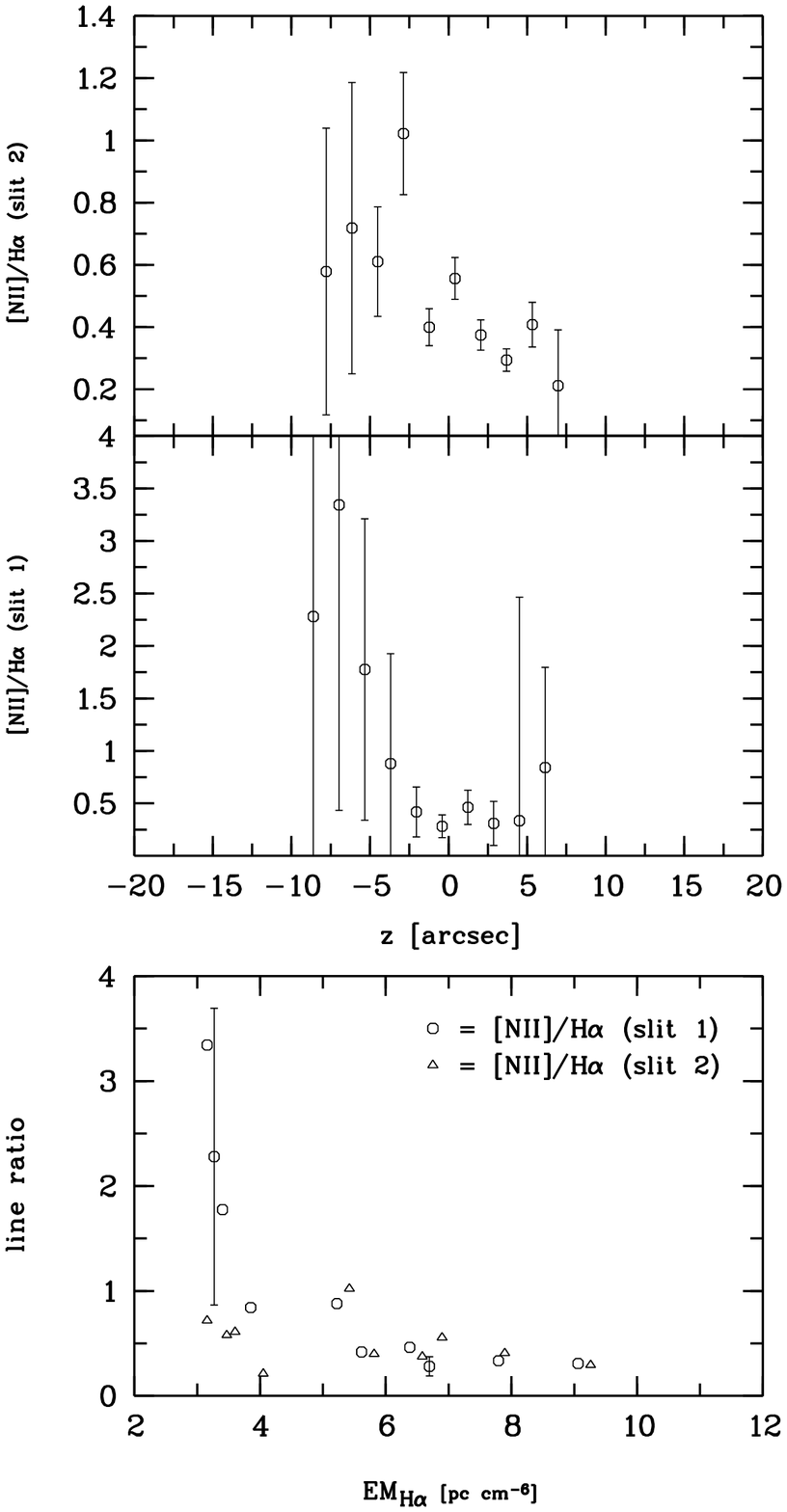,height=9.5cm,width=8.25cm,clip=t,angle=0}
\caption{IC\,2531: Line ratios of [\ion{N}{ii}]\,$\lambda$6583\,/\,H$\alpha$ 
for s1 and s2 (upper panel). Additionally, plots of line ratio vs. 
EM$_{\rm H\alpha}$ together with mean error bars for [\ion{N}{ii}]/H$\alpha$ 
are presented (lower panel). 5$\arcsec$ correspond to 800\,pc.}
\end{figure}

\begin{table}[!h]
\caption{Averaged line ratios for the disk and the halo area of IC\,2531. 
Halo values and their corresponding z-distance are given below. Note that 
the halo data for slit\,1 are only upper limits.}
\label{table3}
\begin{minipage}[t]{8.8cm}
\begin{tabular}{l l l l}
\hline 
\vspace{-0.25cm}
\\

slit 1 & disk & halo (north) & halo (south) \\ 
& & $z=1.1$\,kpc & $z=1.0$\,kpc \\
\vspace{-0.25cm}
\\

\hline
\vspace{-0.2cm}

& & & \\

[\ion{N}{ii}]/${\rm H\alpha}$ & 0.35 $\pm$ 0.17 & 2.02 $\pm$ 2.02 & 0.57 
$\pm$ 1.54 \\
\vspace{-0.2cm}
& & & \\

\hline 
\vspace{-0.25cm}
\\

slit 2 &  disk & halo (north) & halo (south) \\
& & $z=1.0$\,kpc & $z=1.1$\,kpc \\
\vspace{-0.25cm}
\\
\hline
\vspace{-0.2cm}
& & & \\

[\ion{N}{ii}]/${\rm H\alpha}$ & 0.38 $\pm$ 0.07 & 0.80 $\pm$ 0.33 & 0.21 
$\pm$ 0.16\\
\vspace{-0.2cm}
& & & \\

\hline 
\end{tabular} 
\end{minipage}
\end{table}

Because the slits do not cut any bright \ion{H}{ii} region the lack of 
emission lines is not surprising. For both slit positions only the line 
ratios of $\rm{[\ion{N}{ii}]\,\lambda 6585\,/\,H\alpha}$ (Fig.~7) 
could be determined (see also Table~\ref{table3}) and therefore no diagnostic 
diagram could be obtained. 

In view of a single line ratio and relatively large error bars (low S/N ratio)
 it is not 
reasonable to make any statements on possible excitation mechanisms of the 
DIG in IC\,2531. The obtained values for the disk region seem to be 
reproducible by DM94. It is not 
possible to distinguish between a matter or radiation bounded geometry 
because both parameter ranges are able to fit the data.
$\rm{H\alpha}$ line emission can be traced at s1 out to 1.4\,kpc in the 
northern and out to 1.0\,kpc in the southern halo. At s2 the detection
of diffuse  $\rm{H\alpha}$ emission reaches out to 1.3\,kpc in the mean. 

\subsection{NGC\,3044}
\begin{figure*}
\begin{minipage}[ht]{18cm}
\setlength{\unitlength}{1cm}
\begin{picture}(10,8.8)(0,1)
\hspace{0.2cm}
\psfig{file=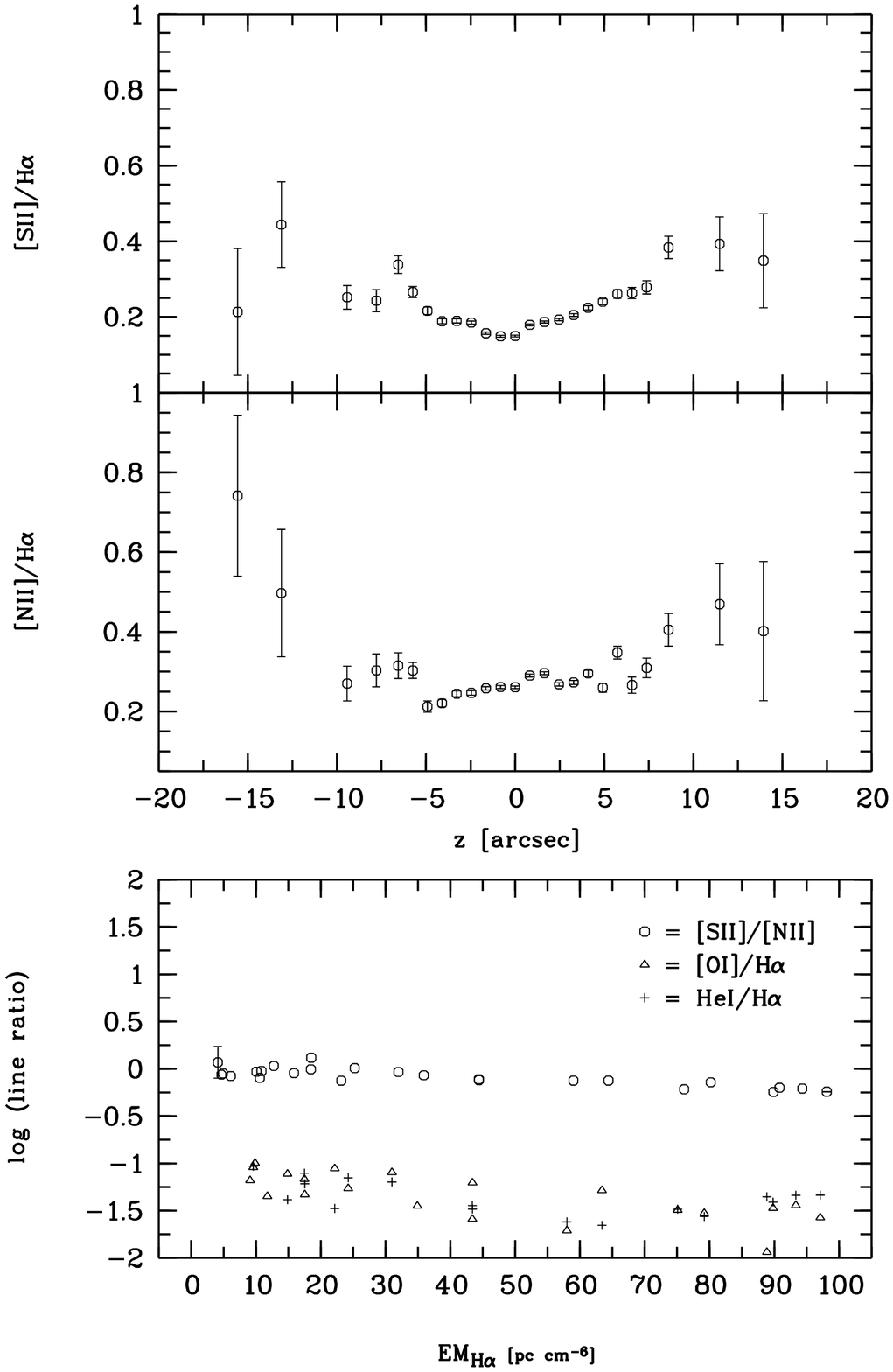,height=9.5cm,width=8cm,clip=t,angle=0}
\hspace{1cm}
\psfig{file=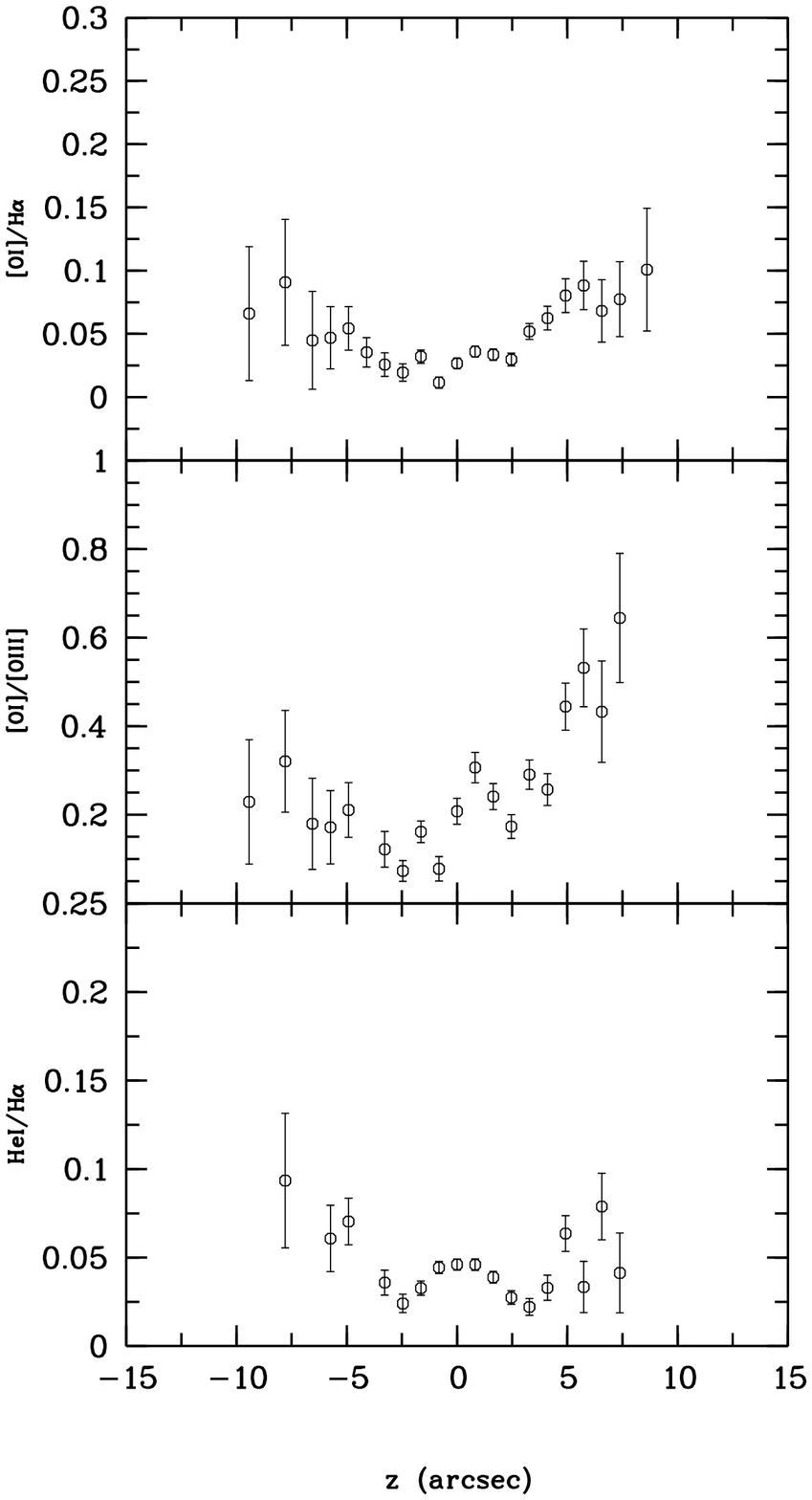,height=9.5cm,width=8cm,clip=t,angle=0}
\end{picture}
\end{minipage}
\vspace{0.5cm}
\\

\parbox[b]{18cm}{\caption{NGC\,3044: Line ratios of 
[\ion{S}{ii}]\,$\lambda$6717\,/\,H$\alpha$,
[\ion{N}{ii}]\,$\lambda$6583\,/\,H$\alpha$  (left) and 
[\ion{O}{i}]\,$\lambda$6300\,/\,H$\alpha$, 
[\ion{O}{i}]\,$\lambda$6300\,/\,[\ion{O}{iii}]\,$\lambda$5007, and 
\ion{He}{i}\,$\lambda$5876\,/\,H$\alpha$ (right) along slit position s1. 
5$\arcsec$ correspond to 415\,pc. The lower left panel contains plots for 
different line ratios vs. EM$_{\rm H\alpha}$, including representative mean 
errors for 
[\ion{S}{ii}]\,/\,[\ion{N}{ii}].}   
\label{fig8}}

\begin{minipage}[hb]{18cm}
\setlength{\unitlength}{1cm}
\begin{picture}(10,8.8)(0,1.75)
\hspace{0.2cm}
\psfig{file=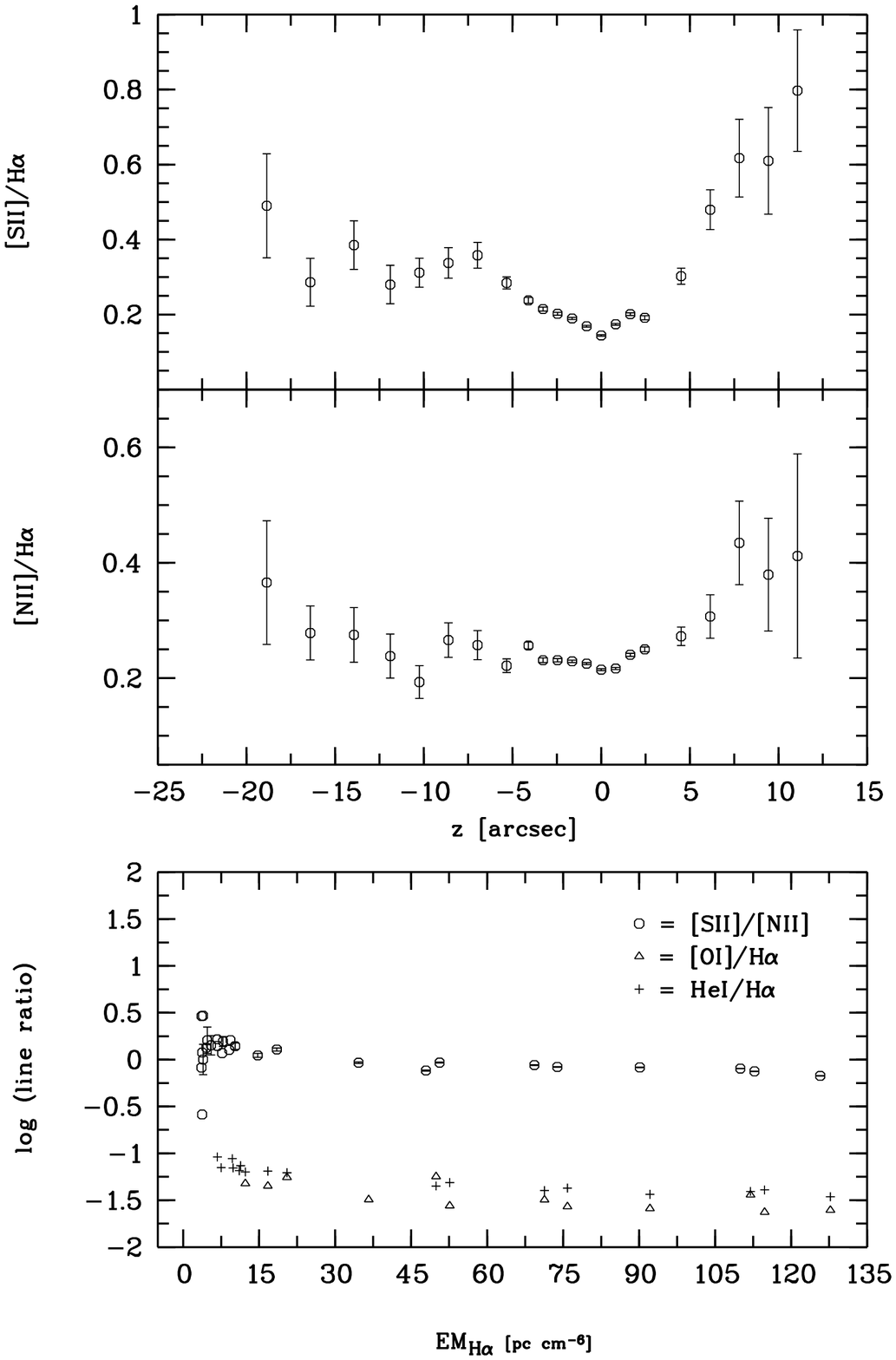,height=9.5cm,width=8cm,clip=t,angle=0}
\hspace{1cm}
\psfig{file=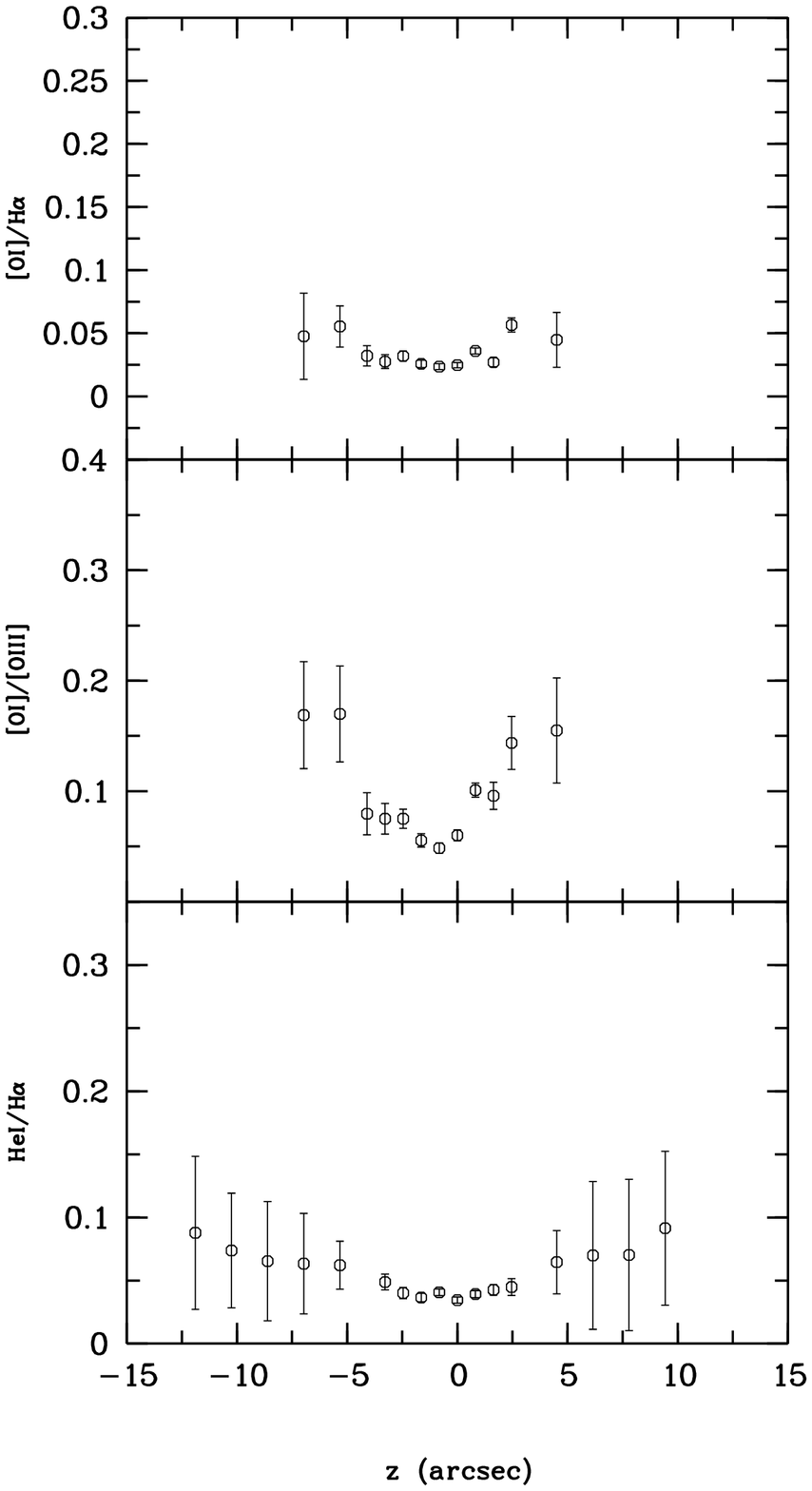,height=9.5cm,width=8cm,clip=t,angle=0}
\end{picture}
\end{minipage}
\vspace{1.25cm}
\\

\parbox[b]{18cm}{\caption{NGC\,3044: Same line ratios of 
[\ion{S}{ii}]\,$\lambda$6717\,/\,H$\alpha$, 
[\ion{N}{ii}]\,$\lambda$6583\,/\,H$\alpha$ (left) and 
[\ion{O}{i}]\,$\lambda$6300\,/\,H$\alpha$, 
[\ion{O}{i}]\,$\lambda$6300\,/\,[\ion{O}{iii}]\,$\lambda$5007, 
 and \ion{He}{i}\,$\lambda$5876\,/\,H$\alpha$ (right) but this time for slit 
position s2. 5$''$ correspond again to 415\,pc. 
Line ratios vs. EM$_{\rm H\alpha}$ are also plotted for s2. Again, note the 
constant behaviour of 
\ion{He}{i}/H$\alpha$ and [\ion{O}{i}]/H$\alpha$ in the lower left panels.}
\label{fig9}}
\end{figure*}

\begin{figure*}
\begin{minipage}[ht]{18cm}
\setlength{\unitlength}{1cm}
\begin{picture}(10,8.8)(0,-2.7)
\hspace{0.35cm}
\psfig{file=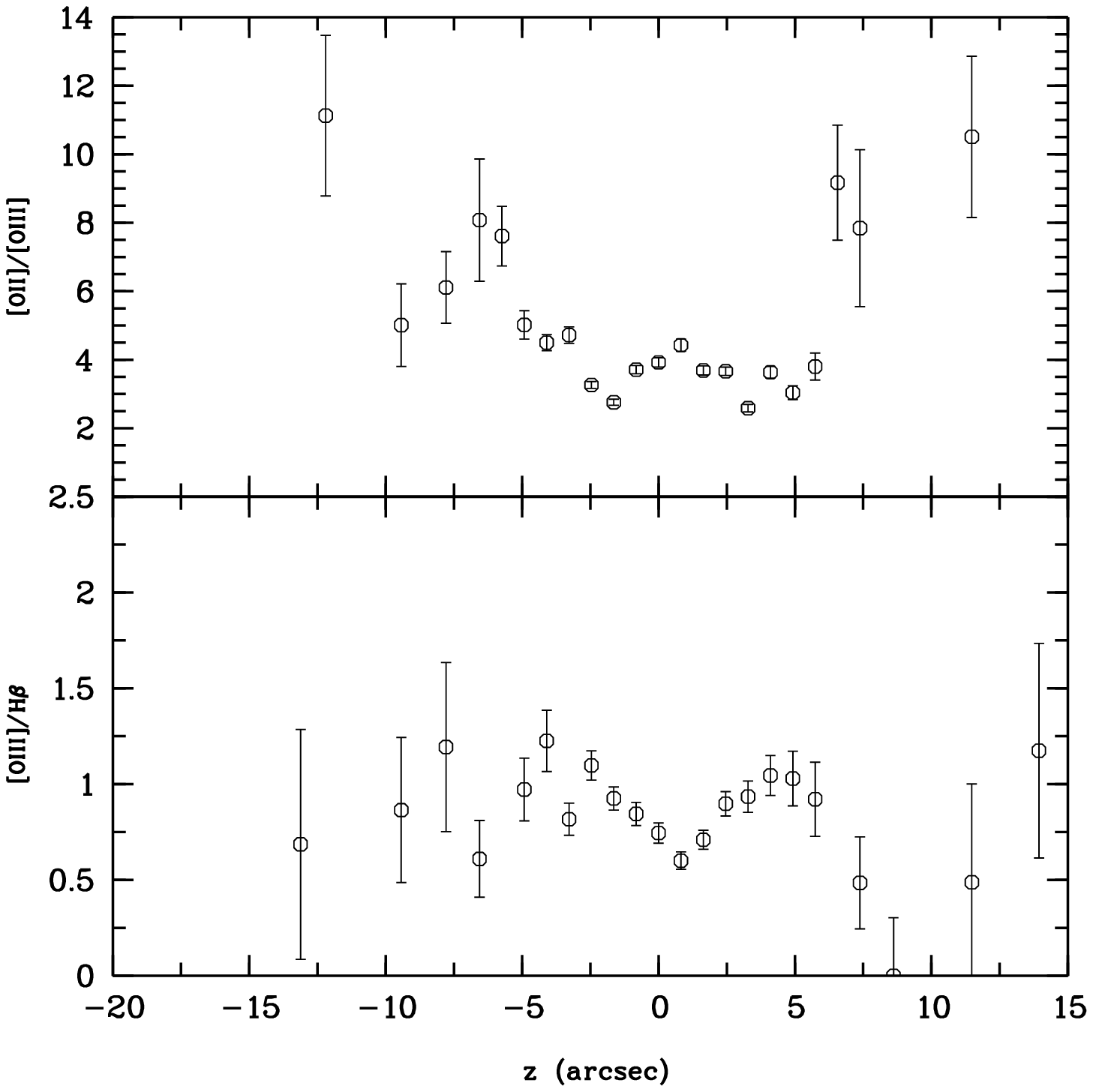,height=6cm,width=7.5cm,clip=t,angle=0}
\hspace{1.6cm}
\psfig{file=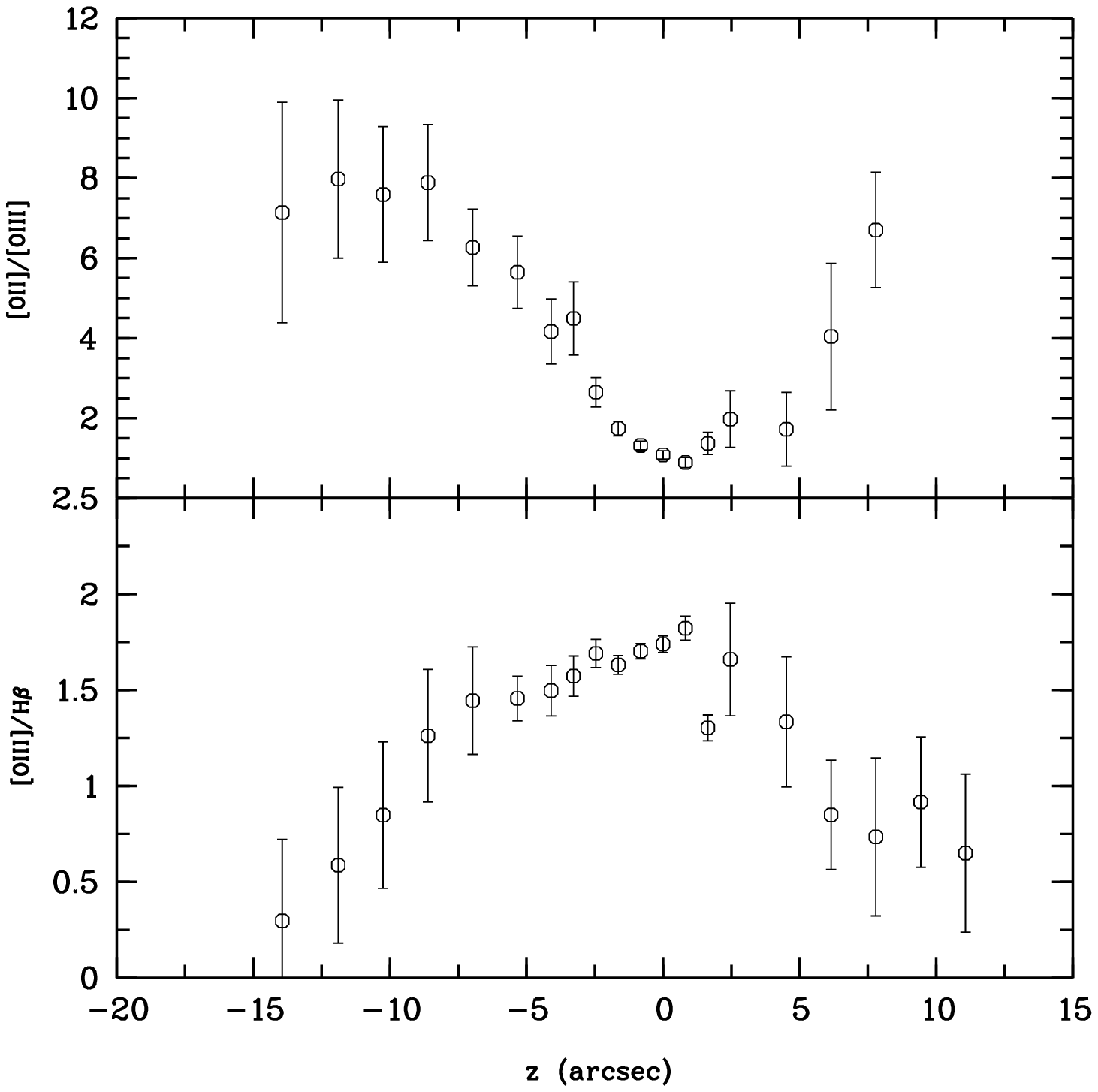,height=6cm,width=7.5cm,clip=t,angle=0}
\end{picture}
\end{minipage}
\vspace{-2.75cm}

\parbox[b]{18cm}{\caption{NGC\,3044: Measured line ratios of 
[\ion{O}{ii}]\,$\lambda$3727\,/\,[\ion{O}{iii}]\,$\lambda$5007 and 
[\ion{O}{iii}]\,$\lambda$5007\,/\,H$\beta$ along slit position s1 (left) and 
s2 
(right) .}
\label{fig10}}

\begin{center}
\hspace{0.0cm}
\psfig{file=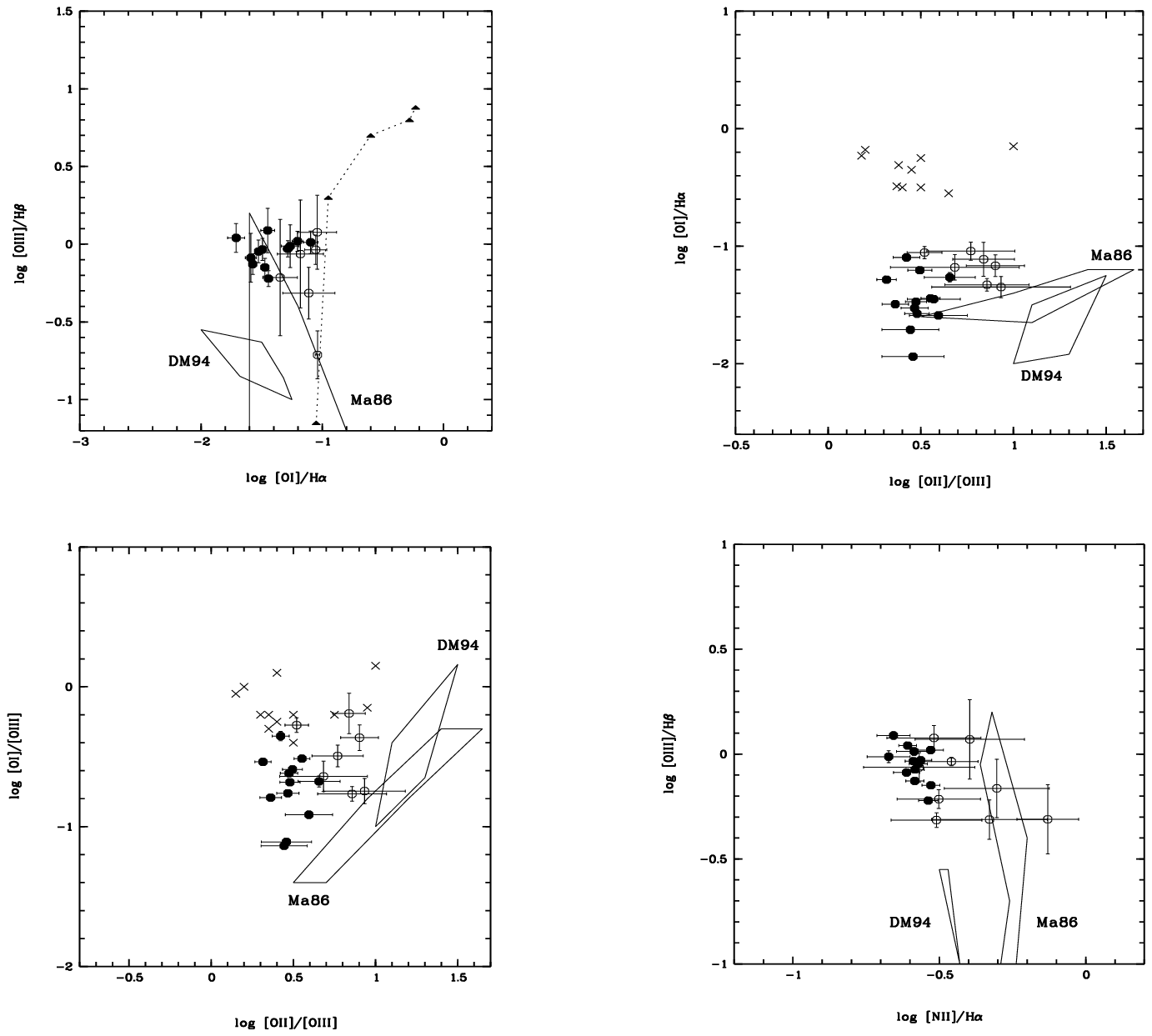,height=13.8cm,width=15cm,angle=0}
\end{center}
\vspace{-0.4cm}
\caption{NGC\,3044: 
Relevant diagnostic diagrams for slit position s1. 
The dotted line and filled triangles are from shock models 
(Shull\,\&\,McKee \cite{shull}) assuming shock velocities between 
80\,--\,100\,${\rm km\ s^{-1}}$. Filled circles denote the disk data 
and open circles address the halo component. Crosses represent 
observed shock ionized objects used by Baldwin et al. (\cite{bald}).
Predicted values of Ma86 and DM92 are enclosed by solid lines.} 
\label{fig11}
\end{figure*}

\begin{table}[t]
\caption{Averaged line ratios for the disk and the halo of NGC\,3044.
Ratios for the halo with an asteriks (*) are measured within $\pm 0.8$\,kpc.}
\label{table4}
\begin{minipage}[t]{8.8cm}
\begin{tabular}{l l l l}
\hline
\vspace{-0.25cm}
\\

slit 1 & disk & halo (north) & halo (south) \\ 
& & $z=1.1$\,kpc & $z=1.2$\,kpc \\
\vspace{-0.25cm}
\\
\hline
\vspace{-0.2cm}

& & & \\

[\ion{N}{ii}]/${\rm H\alpha}$ & 0.25 $\pm$ 0.01 & 0.51 $\pm$ 0.16 & 0.37 
$\pm$ 0.06\\ 

[\ion{S}{ii}]/${\rm H\alpha}$ & 0.19 $\pm$ 0.04 & 0.32 $\pm$ 0.14 & 0.33 
$\pm$ 0.04 \\
 
\ion{He}{i}/${\rm H\alpha}^{*}$ & 0.04 $\pm$ 0.01 & 0.08 $\pm$ 0.03 & 0.06 $\pm$ 
0.02  \\

[\ion{O}{i}]/[\ion{O}{iii}]$^{*}$ & 0.17 $\pm$ 0.03 & 0.27 $\pm$ 0.09 & 0.54 $\pm$ 
0.13 \\ 

[\ion{O}{i}]/${\rm H\alpha}^{*}$ & 0.02 $\pm$ 0.01 & 0.07 $\pm$ 0.04 & 0.08 
$\pm$ 0.04 \\ 

[\ion{O}{iii}]/${\rm H\beta}$ & 0.91 $\pm$ 0.10 & 0.74 $\pm$ 0.29 & 0.49 
$\pm$ 0.38\\ 

[\ion{O}{ii}]/[\ion{O}{iii}] & 2.92 $\pm$ 0.27 & 6.94 $\pm$ 2.02 & 6.11 $\pm$ 
1.50 \\ 
\vspace{-0.2cm}
& & & \\

\hline
\vspace{-0.25cm}
\\
 slit 2 &  disk  &  halo (north) &  halo (south) \\
 & & $z=1.2$\,kpc & $z=0.9$\,kpc \\
\vspace{-0.25cm}
\\
\hline
\vspace{-0.2cm}
& & & \\

[\ion{N}{ii}]/${\rm H\alpha}$ & 0.24 $\pm$ 0.01 & 0.28 $\pm$ 0.07 & 0.35 
$\pm$ 0.04\\
 
[\ion{S}{ii}]/${\rm H\alpha}$ & 0.19 $\pm$ 0.01 & 0.41 $\pm$ 0.09 & 0.64 
$\pm$ 0.11\\

\ion{He}{i}/${\rm H\alpha}^{*}$ & 0.04 $\pm$ 0.01 & 0.07 $\pm$ 0.04 & 0.08 
$\pm$ 0.06 \\

[\ion{O}{i}]/[\ion{O}{iii}]$^{*}$ & 0.06 $\pm$ 0.03 & 0.17 $\pm$ 0.04 & \mbox{
\hspace{0.55cm}} --- \\ 

[\ion{O}{i}]/${\rm H\alpha}^{*}$ & 0.04 $\pm$ 0.01 & 0.05 $\pm$ 0.03 & \mbox{
\hspace{0.55cm}} --- \\ 

[\ion{O}{iii}]/${\rm H\beta}$ & 1.66 $\pm$ 0.10 & 0.87 $\pm$ 0.27 & 1.15 
$\pm$ 0.37\\

[\ion{O}{ii}]/[\ion{O}{iii}] & 2.45 $\pm$ 0.42 & 6.39 $\pm$ 1.38 & 5.73 $\pm$ 
1.56 \\ 
\vspace{-0.2cm}
& & & \\

\hline
\end{tabular}
\end{minipage}
\end{table}

Besides NGC\,1963 this galaxy shows the strongest diffuse extraplanar emission.
 Again, an $\rm{H\alpha}$ image from Rossa\,\&\,Dettmar was used to classify 
the DIG morphology as ``bright'' and ``diffuse'' (cf. Fig.~\ref{fig1}). 
For NGC\,3044 line ratios are presented in Figs.~\ref{fig8}\,--\,\ref{fig10}, 
and diagnostic diagrams are shown in Fig.~\ref{fig11}. Averaged line ratios 
are given in Table~\ref{table4}. 
In this galaxy diffuse extraplanar H$\alpha$ emission is detectable at slit 
s1 up to $\pm$\,15\arcsec\ which corresponds to 1.3\,kpc. At s2 the 
extraplanar DIG can 
be traced up to 1.6\,kpc in the northern and up to 1.1\,kpc in southern halo.
All line ratios except that of [\ion{S}{ii}]/H$\alpha$ at slit s2 for 
NGC\,3044 are well reproduced by the photoionization model from Ma86 or 
DM94 assuming a matter 
bounded case with $\rm{X_{edge} = 0.10}$ and $\rm{q= 0.001}$ ($q$ is 
proportional to the ratio of ionizing photon density to electron density 
cubed and also proportional to the ionization parameter $U$). In these models 
ionizing sources are O5 stars with temperatures of 
$\rm{T_{*} = 4.5\cdot 10^{4}~K}$. 
This means that most of the hard Lyman continuum radiation from the star 
forming regions can escape and ionize the medium at high 
galactic latitudes. 
The above mentioned line ratio [\ion{S}{ii}]/H$\alpha$ reaches values of 
0.80\,$\pm$\,0.16 at z\,$=$\,11\arcsec. This finding is in good agreement with
 the data obtained by Lehnert\,\&\,Heckman (\cite{lehe}). 
In diagnostic diagrams including the 
[\ion{O}{i}] emission line (Fig.~11) the data for the halo
fall into the gap between \ion{H}{ii} regions and shock ionized objects. 
Additionally, the separation of disk and halo components is not as clear as 
in NGC\,1963. 
The line ratio of [\ion{S}{ii}]/H$\alpha$ for the southern halo at s2, the 
position of our data in diagnostic diagrams (Fig.~11), and a moderate gradient
 in line ratios reveal the hybrid character of NGC\,3044 regarding the
excitation mechanism. Although photoionization seems to be the main
ionizing source in this galaxy contributions due to shocks cannot be ruled out.
Out to our detection limit we find again a decreasing
$\rm{[\ion{O}{iii}]/H\beta}$ ratio with
increasing $|z|$.

\subsection{NGC\,4302}
No diffuse extraplanar line radiation is detectable in NGC\,4302 at slit 
position s1 above
 $2\sigma$.  
Slit s2 cuts the outer south western part of an extended \ion{H}{ii} region.
Therefore only faint $\rm{H\alpha}$ and [\ion{N}{ii}]\,$\lambda$6583
 emission from the disk can be detected.
No empirical diagnostic diagram could be obtained. Averaged values for
 [\ion{N}{ii}]\,$\lambda$6583/$\rm{H\alpha}$ are shown for ${\rm
 -4\arcsec \le |z| \le +4\arcsec}$ in Table~\ref{table5}, with
 4\arcsec\ corresponding to 360\,pc.
The DM94 model fails in predicting the measured
data while the "simplified" Ma86 model fits with
respect to the error bars values up to 0.62 well. This assumes a
diluted radiation field ($\rm{\log q} = -4$) and O5 stars ($\rm{T_{*}
= 4.5\cdot 10^{4}~K}$) as ionizing sources. For values larger than
0.62, as observed in the outer disk, the Mathis models fail, too. 

\begin{table}[h]
\caption{Averaged line ratios for the center, the 
northern (N), and southern (S) outer disk of NGC\,4302 at slit position s2.}
\label{table5}
\begin{minipage}[t]{8.8cm}
\begin{tabular}{l l l l}
\hline
\vspace{-0.25cm}
\\
 slit 2 &  disk center &  outer disk (N) & outer disk (S) \\ 
\vspace{-0.25cm}
\\
\hline
\vspace{-0.2cm}
& & & \\
\vspace{-0.2cm}

[\ion{N}{ii}]/${\rm H\alpha}$ & 0.44 $\pm$ 0.06 & 0.78 $\pm$ 0.24 & 0.99 
$\pm$ 0.24\\

& & & \\

\hline 
\end{tabular} 
\end{minipage}
\end{table}

\subsection{NGC\,4402}
An R-band image of NGC\,4402 reveals that this galaxy is rich in dust and 
shows some spectacular dust filaments emerging from the south eastern and 
western parts of the disk.
The corresponding $\rm{H\alpha}$ image gives evidence that these filaments 
are connected with \ion{H}{ii} regions (star forming regions) inside the disk. 
Our slit position has been chosen such that the star in the north of the 
galaxy falls onto the slit (Fig.~\ref{fig1}).
Despite of a ``diffuse'' DIG classification no emission lines in the blue 
wavelength domain could be detected.
The measured line ratios are presented in Fig.~12 and 
averaged values are given in Table~\ref{table6}. 

\begin{table}[h]
\caption{Averaged line ratios for the disk and halo area of NGC\,4402.}
\begin{minipage}[t]{8.8cm}
\begin{tabular}{l l l l}
\hline 
\vspace{-0.25cm}
\\

slit 1 & disk & halo (north) & halo (south) \\ 
 & & $z=1.9$\,kpc & $z=1.3$\,kpc \\
\vspace{-0.25cm}
\\

\hline
\vspace{-0.2cm}
& & & \\

[\ion{N}{ii}]/${\rm H\alpha}$ & 0.30 $\pm$ 0.05 & 0.76 $\pm$ 0.34 & 1.07 
$\pm$ 0.29 \\

[\ion{S}{ii}]/${\rm H\alpha}$ & 0.15 $\pm$ 0.04 & 0.50 $\pm$ 0.24 & 0.62 
$\pm$ 0.18\\
\vspace{-0.2cm}
& & & \\

\hline 
\end{tabular} 
\end{minipage}
\label{table6}
\end{table}

For the disk region of NGC\,4402 the DM94 model reproduces the observed 
ratios well. 
In order to fit the data one has to choose a matter bounded model 
($\rm{X_{edge} = 0.10}$) and a photon field with $\rm{\log q = -3}$. 
Roughly 20\,\% of the ionizing photons escape from the \ion{H}{ii} regions 
and are able to ionize sulfur and nitrogen in the halo. 
Higher values ($<$ 0.9) for the halo are reproduced (with respect to the 
error bars) by the model of Ma86 using O5 stars as 
ionizing sources and a still softer radiation field ($\rm{\log q = -4}$). 
Values of 0.62 for [\ion{N}{ii}]/H$\alpha$ and 0.61 for 
[\ion{S}{ii}]/H$\alpha$ are predicted. Ratios exceeding 0.9 lead to a model 
failure.
  
\begin{figure}[t]
\hspace{0.25cm}
\psfig{file=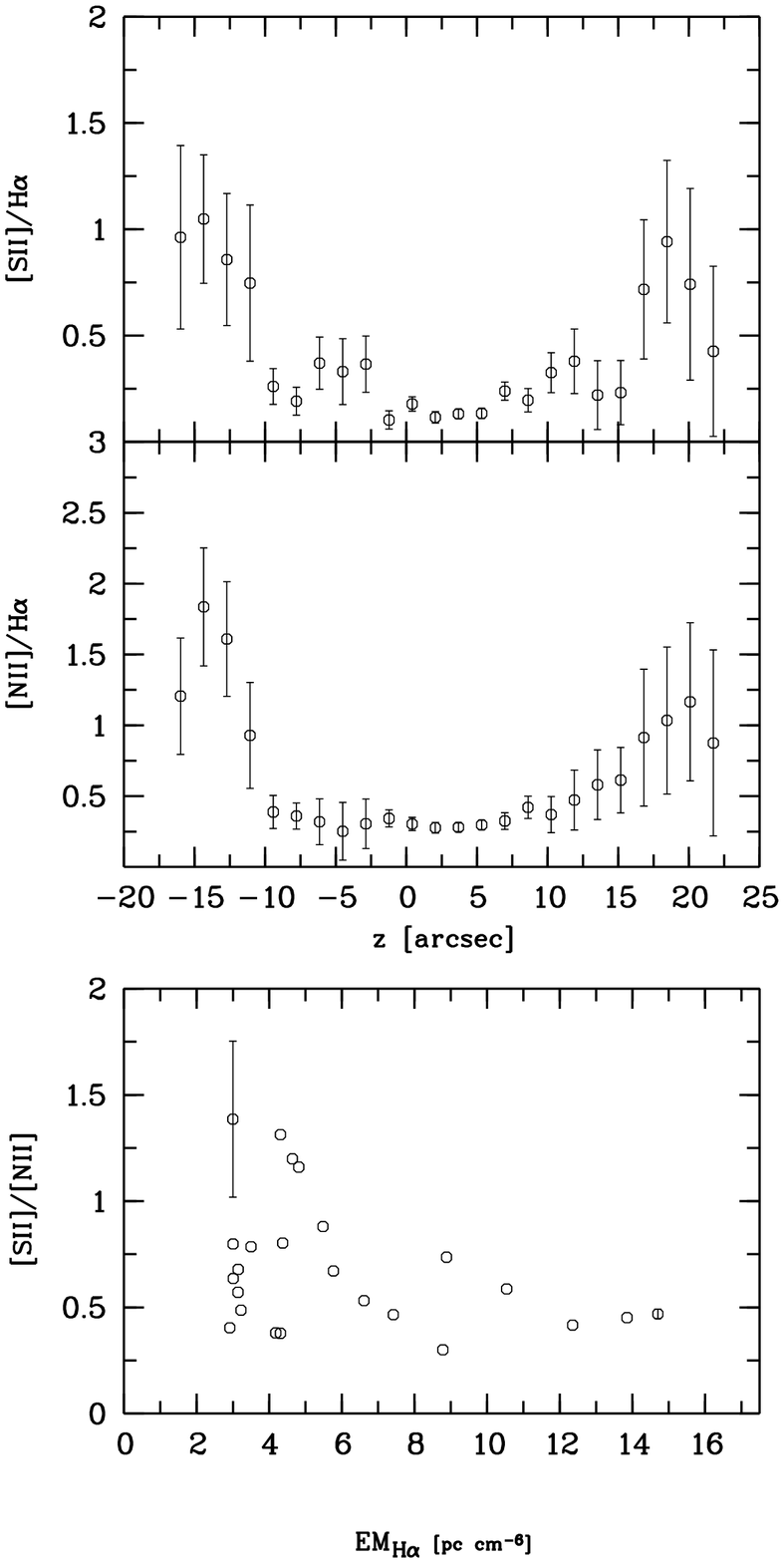,height=9.5cm,width=8cm,clip=t,angle=0}
\caption{NGC\,4402: Line ratios of 
[\ion{S}{ii}]\,$\lambda$6717\,/\,H$\alpha$ and 
[\ion{N}{ii}]\,$\lambda$6583\,/\,H$\alpha$ including variations of 
[\ion{S}{ii}]/[\ion{N}{ii}] vs. EM$_{\rm H\alpha}$ 
for the only slit position s1. 5$\arcsec$ correspond to 530\,pc.
Positive $z$-values 
denote the northern, and negative the southern halo.}
\end{figure}

The steep gradient in [\ion{N}{ii}]/H$\alpha$ or [\ion{S}{ii}]/H$\alpha$ 
reaching values of 1 or higher at high $|z|$ cannot be explained in
the framework of dilute photoionization models and may indicate again 
the need for an additional heating source.

\subsection{NGC\,4634}
 On the $\rm{H\alpha} + [\ion{N}{ii}]$ image for NGC\,4634 a bright and 
extraplanar DIG layer is visible. Additionally, several filaments emerge 
mainly from the north-eastern part of the disk (see Fig.~\ref{fig1}). Here 
again the slits had been positioned such that they contain stars for
accurate positioning. 
Slit  s1 cuts the brigthest \ion{H}{ii} region $\approx$ 5\arcsec\ north 
west of an extended dust cloud and s2 covers an area of fainter 
$\rm{H\alpha}$ intensity $\approx$\,14\arcsec\ north west of s1. 
As in NGC\,4402, only the spectral lines of [\ion{N}{ii}]\,$\lambda
\lambda$6549, 6583, $\rm{H\alpha}$, and [\ion{S}{ii}]\,$\lambda\lambda$6717,
6732 are measurable at both slit positions above $2\sigma$.
The corresponding line ratios are plotted in Figs.~13 and 14. 
Due to the non-detection of 
emission lines in the blue wavelength region no diagnostic diagrams 
could be created.
In NGC\,4634 DIG emission can be traced up to 1.2\,kpc at slit position s1 
and up to 0.9\,kpc at s2. 
Measured line ratios of [\ion{N}{ii}]/H$\alpha$ for the disk area can be 
reproduced using the DM94 model and setting 
$\rm{X_{edge}}$ again to 0.10 and $\rm{\log q}$ to $-4$. Only 10\,\% of 
hydrogen is neutral at the border $\rm{H^{0},H^{+}}$ that is 31\,\% of the 
ionizing photons can escape the geometry. The model fails again for values 
larger than 0.37. [\ion{S}{ii}]/H$\alpha$ values of the order 
0.20\,$\pm$\,0.02 (cf. Table~\ref{table7}) are also not predicted by this 
geometry. 

\begin{figure}[!pt]
\hspace{0.25cm}
\psfig{file=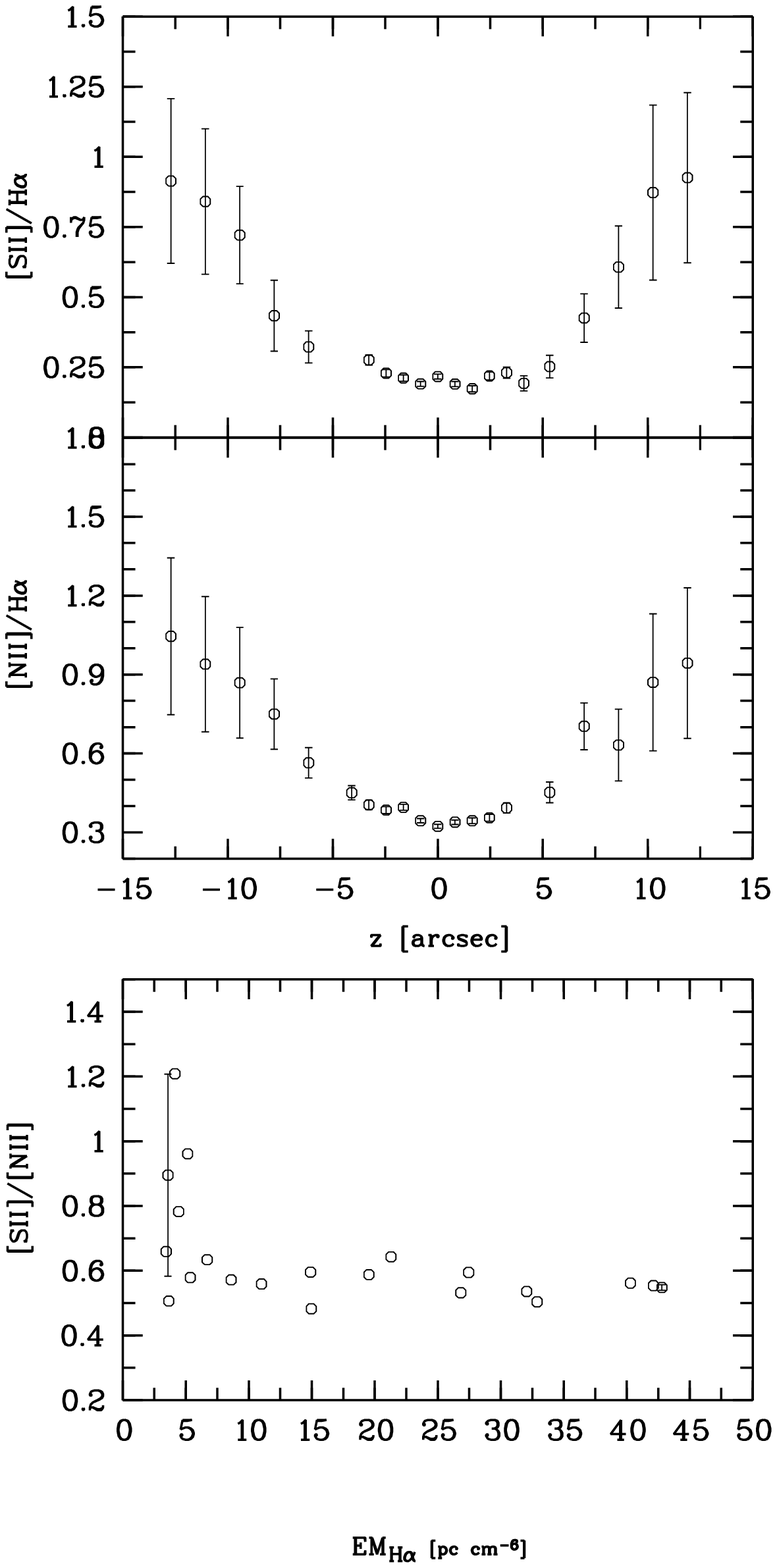,height=9.5cm,width=8cm,clip=t,angle=0}
\caption{NGC\,4634: Line ratios of [\ion{S}{ii}]\,$\lambda$6717\,/\,H$\alpha$ 
and [\ion{N}{ii}]\,$\lambda$6583\,/\,H$\alpha$ for slit s1. 5$''$ correspond 
to 465\,pc. The lower panel displays 
[\ion{S}{ii}]/[\ion{N}{ii}] vs. EM$_{\rm H\alpha}$ in combination with 
corresponding mean errors.}
\end{figure}

\begin{figure}[!pt]
\hspace{0.25cm}
\psfig{file=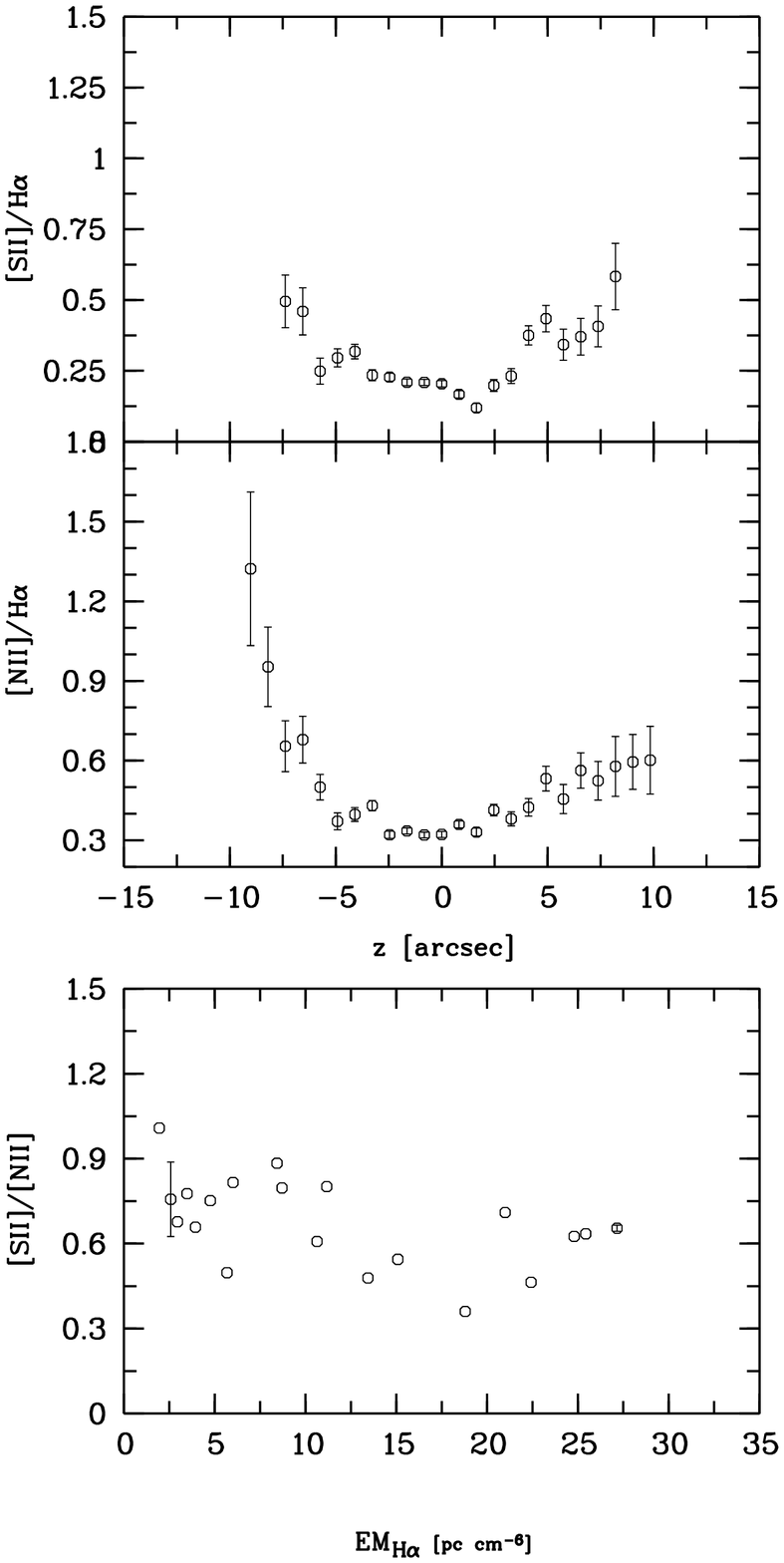,height=9.5cm,width=8cm,clip=t,angle=0}
\caption{NGC\,4634: Line ratios of [\ion{S}{ii}]\,$\lambda$6717\,/\,H$\alpha$ 
and [\ion{N}{ii}]\,$\lambda$6583\,/\,H$\alpha$ for slit s2. 5$''$ correspond 
to 465\,pc. The lower panel shows again 
[\ion{S}{ii}]/[\ion{N}{ii}] vs. EM$_{\rm H\alpha}$ 
along with the representative errors.}
\end{figure}

\begin{table}[t]
\caption{Averaged line ratios for the disk and halo area of NGC\,4634.}
\label{table7}
\begin{minipage}[t]{8.8cm}
\begin{tabular}{l l l l}
\hline 
\vspace{-0.25cm}
\\

slit 1 & disk & halo (north) & halo (south) \\
& & $z=1.2$\,kpc & $z=1.2$\,kpc \\
\vspace{-0.25cm}
\\

\hline
\vspace{-0.2cm}
& & & \\

[\ion{N}{ii}]/${\rm H\alpha}$ & 0.39 $\pm$ 0.02 & 0.79 $\pm$ 0.21 & 0.80 
$\pm$ 0.18 \\

[\ion{S}{ii}]/${\rm H\alpha}$ & 0.22 $\pm$ 0.02 & 0.69 $\pm$ 0.19 & 0.62 
$\pm$ 0.18\\
\vspace{-0.2cm}
& & & \\

\hline 
\vspace{-0.25cm}
\\

slit 2 & disk & halo (north) & halo (south) \\ 
& & $z=0.8$\,kpc & $z=0.8$\,kpc \\
\vspace{-0.25cm}
\\

\hline
\vspace{-0.2cm}
& & & \\

[\ion{N}{ii}]/${\rm H\alpha}$ & 0.34 $\pm$ 0.01 & 0.50 $\pm$ 0.07 & 0.85 
$\pm$ 0.16 \\

[\ion{S}{ii}]/${\rm H\alpha}$ & 0.18 $\pm$ 0.02 & 0.51 $\pm$ 0.08 & 0.37 
$\pm$ 0.07\\
\vspace{-0.2cm}
& & & \\

\hline 
\end{tabular} 
\end{minipage}
\end{table}

Additional information on ionization sources and resulting line ratios can 
only be made by using the simplified Ma86 model. 
Stars of spectral type O5 ($\rm{T_{*} = 4.5\cdot 10^{4}~K}$) are taken to be
responsible for the ionization of the observed elements. 
Taking error bars into account halo values for [\ion{N}{ii}]/H$\alpha$ 
are well reproduced, as long as they are $\le$\,0.62. Larger ratios lead to a 
failure of the model.
Relative line strenghts of [\ion{S}{ii}]/H$\alpha$ (s1 and s2) are well 
fitted by Ma86. It produces values of 0.62 which is consistent 
with the measured data especially for slit position s1.
Although precise statements on excitation mechanisms of the extraplanar DIG 
cannot be made with the present data, the failure of the photoionization 
models implies more than one ionizing/heating source.

\subsection{Discussion}
Recent studies of DIG in the Milky Way  and several edge-on galaxies
confirm a nearly constant trend for [\ion{S}{ii}]/H$\alpha$ vs. 
[\ion{N}{ii}]/H$\alpha$ (e.g., Haffner et al. \cite{haff}) and also for 
[\ion{S}{ii}]/[\ion{N}{ii}] vs. $z$ (Rand \cite{rand}).
These observations cannot be reproduced by photoionization  
models which depend on the ionization parameter $U$, because as  
[\ion{N}{ii}]/H$\alpha$ and [\ion{S}{ii}]/H$\alpha$ increase towards the halo,
due to a diluted radiation field (smaller $U$), [\ion{S}{ii}]/[\ion{N}{ii}] 
increases, too.
To explain this finding Reynolds et al. (\cite{rey99}) recently
proposed an additional heating 
source that is proportional to $n_{e^{-}}$ and increases the electron 
temperature towards the outer halo.

\addtocounter{figure}{0}
\begin{figure}[!t]
\begin{center}
\psfig{file=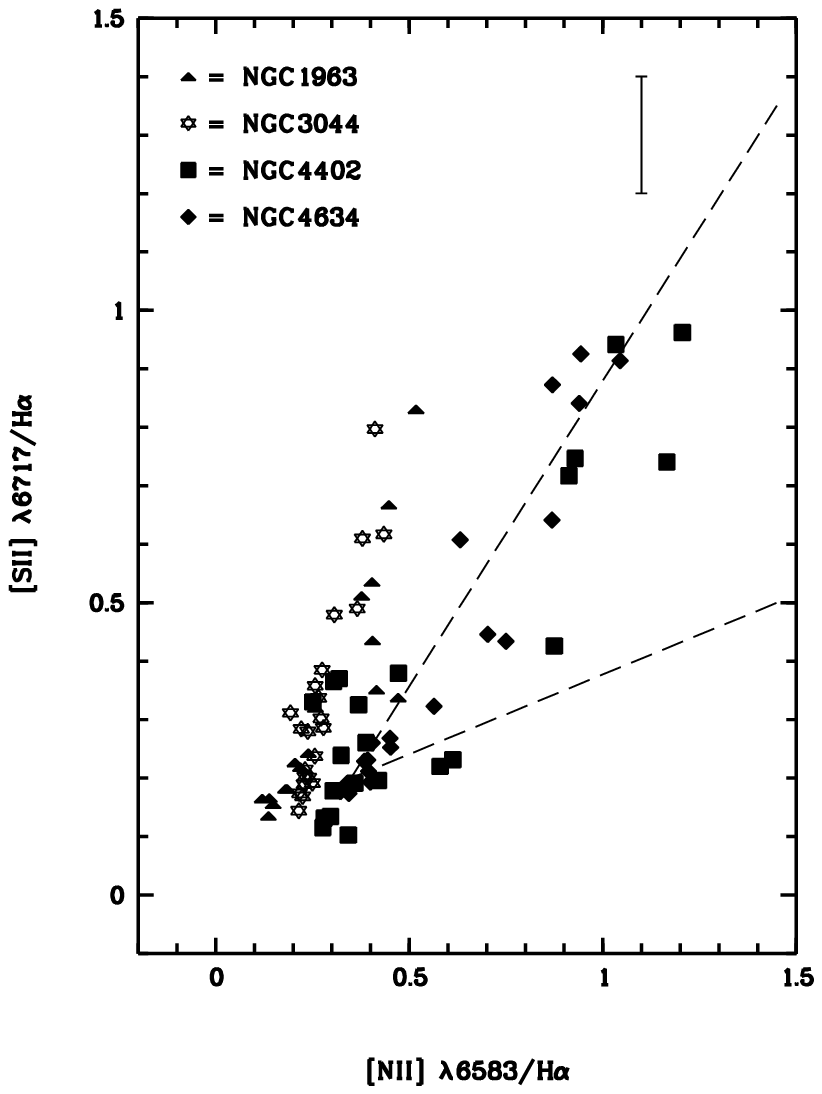,height=7.6cm,width=8cm,angle=0}
\caption{[\ion{S}{ii}]\,$\lambda$6717/H$\alpha$ vs.
[\ion{N}{ii}]\,$\lambda 6583$/H$\alpha$ 
for all galaxies with nitrogen and sulfur emission. The mean error for the 
halo of all 
galaxies is plotted in the upper right corner. For comparison, data for the 
Milky Way
 (Haffner et al. \cite{haff}) and NGC\,891 (Rand \cite{rand}) would be located 
between the dashed lines.}
\end{center}
\label{fig13}
\end{figure}
\addtocounter{figure}{0}
\begin{figure}[!ht]
\begin{center}
\psfig{file=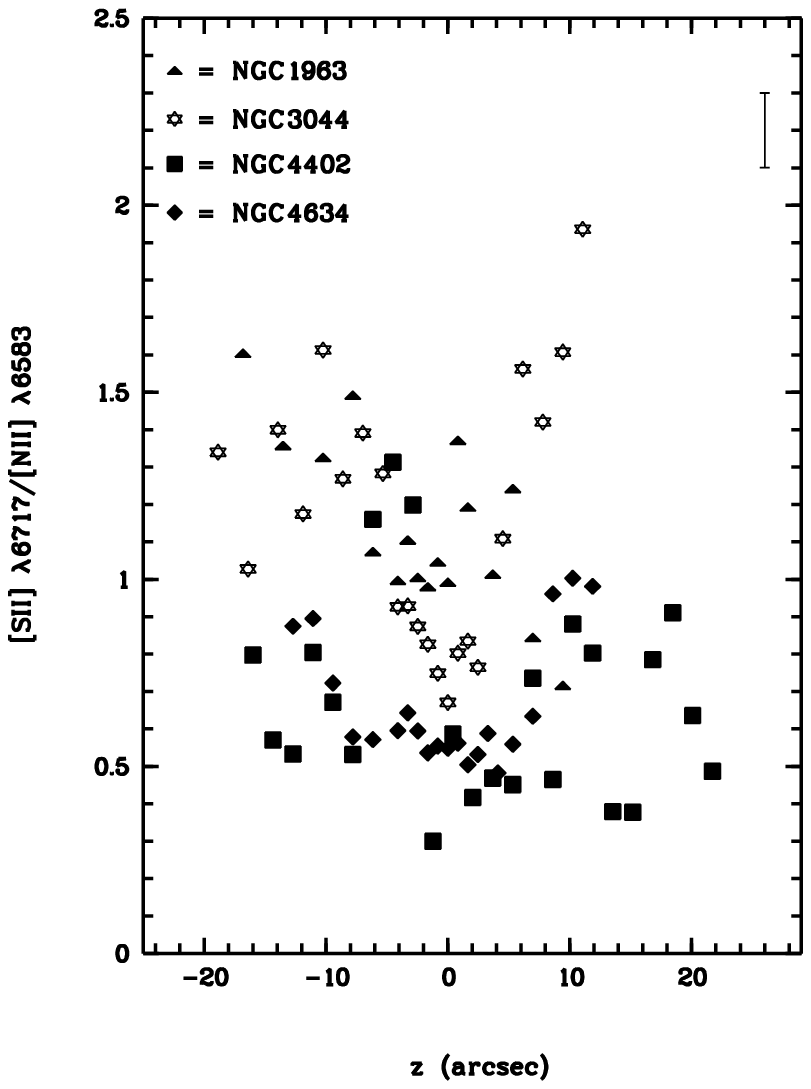,height=7.6cm,width=8cm,angle=0}
\caption{Alternative plot of 
[\ion{S}{ii}]\,$\lambda$6717/[\ion{N}{ii}]\,$\lambda 6583$ vs. distance $z$
perpendicular to the plane. Now, different gradients are visible in both halo 
hemispheres.}
\end{center}
\label{fig14}
\end{figure}

Ideas for possible heating processes reach from turbulent dissipation to
magnetic reconnection. In the low density environment these processes 
would be more efficient than photo\-ioniza\-tion, 
thus providing the necessary heating rate to account for the shapes of the 
above mentioned line ratios and possibly also for the rise of
[\ion{O}{iii}]/H$\beta$ with z (Reynolds et al. \cite{rey99}).
We present in Fig.~15 plots of 
[\ion{S}{ii}]/H$\alpha$ vs. [\ion{N}{ii}]/H$\alpha$ supporting a linear 
dependence. Since line ratios are given with respect to H$\alpha$
intensities, this relation holds with respect to densities.  If 
 [\ion{S}{ii}]/[\ion{N}{ii}] line ratios are instead plotted vs. the 
geometrical unit z (Fig.~16) the scatter is much increased. 
In this presentation significant non-linear changes in 
[\ion{S}{ii}]/[\ion{N}{ii}] vs. z can be noticed. 
In some cases (NGC\,3044, NGC\,1963 and NGC\,4634) the halo
hemispheres may have different slopes which could be explained by, e.g., 
different metallicities (Haffner et al. \cite{haff}).
Plots of line ratios vs. emission measure EM$_{H\alpha}$ as a first estimate
of a density dependence (lower panels of Figs.~4,9,12, and 14) also
support a non-linear trend. 

As can be seen from Fig.~16 the ratio of [\ion{S}{ii}]/[\ion{N}{ii}] for 
NGC\,1963 increases
 rapidly from 1.0 in the disk at $z=0''$ to 1.6 at $z=-18''$ towards the halo.
Values of 1.12\,--\,1.34 are predicted by DM94 assuming the same parameter 
setting as 
mentioned in Sect.~{\it 4.1} reflecting the observed data for the disk and 
the northern halo region well. For ratios less than 1 the chosen parameter 
set which previously reproduced [\ion{N}{ii}]/$\rm{H\alpha}$ and 
[\ion{S}{ii}]/$\rm{H\alpha}$ now leads to a model failure 
(southern halo, cf. Fig.~16). Alternative settings are also unable to
fit observations.
A likely physical reason could be a small scale fluctuation (increase) 
of the electron temperature $T_{e}$ due to a decrease in metallicity or 
density of the extraplanar DIG.  Although the model 
cannot account for local density or metallicity variations the general trend 
of [\ion{N}{ii}]/$\rm{H\alpha}$ or [\ion{S}{ii}]/[\ion{N}{ii}] is 
qualitatively well reproduced.

Compared to NGC\,1963 the shape of [\ion{S}{ii}]/[\ion{N}{ii}] for NGC\,3044 
 looks similar but reveals a lower starting value (0.67) and a non-linear 
gradient for $|z| < -10\arcsec$  (compare to Fig.~15 where the 
gradient seems to be linear). With respect to the mean errors, the 
observed ratios are again best represented by the DM94 model. 
 The scatter in [\ion{S}{ii}]/[\ion{N}{ii}] (Fig.~16) at 
$-20\arcsec<z<-10\arcsec$ 
suggests that the electron temperature and most likely also the metallicity
is a function of z, varying significantly on large scales ($\approx$ 1 kpc). 
Within our sample NGC\,1963 and NGC\,3044 are showing the steepest gradients 
and 
largests mean values of [\ion{S}{ii}]/[\ion{N}{ii}], hence deviating from 
Rand's observation in NGC\,891. 

The ratio of [\ion{S}{ii}]/[\ion{N}{ii}] for NGC\,4402 (Fig.~15 or 16) 
reveals
 only a relatively flat gradient towards the halo. This result differs 
extremely from the one for NGC\,1963 or NGC\,3044, and is similar to the 
ratio found for NGC\,891. It also indicates that none of our selected models 
can predict 
correct values, especially not for the most extreme points in this plot.
It could be argued that diluted and very soft 
radiation is responsible for the observed values. 
However, this is not probable since [\ion{N}{ii}] should be increased, too.

Haffner et al. (\cite{haff}) have shown for the Milky Way that different 
slopes in [\ion{S}{ii}]/H$\alpha$ vs. [\ion{N}{ii}]/H$\alpha$ could be an 
indicator for different metallicities. As a result the individual gradients 
visible in Fig.~16 could indicate different metallicities for each halo 
hemisphere. If this is true NGC\,4402 would have the same metallicity as the 
Milky Way or its "twin" NGC\,891.           

The data for NGC\,4634 in Fig.~16 are in agreement with Ma86 
and demonstrate a constant run at 0.6 for the disk, followed by a relatively 
steep gradient. At $z=\pm 10\arcsec$ the 
ratio appears to remain constant at 1.0. 
In the disk [\ion{S}{ii}]/[\ion{N}{ii}] is 
comparable to that of NGC\,891 but with increasing $|z|$ this similarity 
fades.

In summary we note that strong nitrogen lines ($> 0.62$ with
respect to hydrogen) are generally not well reproduced by pure photoionization
models. The general trend of [\ion{S}{ii}]/[\ion{N}{ii}] is 
reproduced qualitatively but not quantitatively (models cannot account for 
localized changes).
They even fail in fitting oxygen and helium line ratios for 
the halo. The diagnostic diagram using [OIII] and [OI] for NGC\,1963 indicates
that shocks may contribute as an ionizing and heating source in the halo. 
This is similar to the finding of Martin (\cite{martin}) for diffuse
ionized gas in the outflows of dwarf galaxies. 
However, for the objects studied here, the discussed line ratios do not 
only change with geometrical distance from the midplane of the disk, i.e. 
with respect to the 
location of the suspected ionizing sources. We rather find sudden localized
changes, most evidently in NGC\,3044 and NGC\,4402, indicating that
ionization conditions and/or heating rates may change on length scales of a
few hundred parsecs. 
These non-linear changes can also be seen in the correlation between 
[SII]/[NII] line ratios and emission measure (Figs.~4,9,12, or 14). 
Since they also correlate with the local density of the diffuse
medium, these variations can only be explained by small scale density 
fluctuations.

\subsection{Kinematics}
In order to probe our results and to establish a relationship between 
excitation mechanisms and gas kinematics
 we have plotted in Fig.~17 an additional diagnostic diagram regarding 
line width (FWHM) vs. line ratio following, e.g., 
Lehnert\,\&\,Heckman (\cite{leheck}). 
We have chosen [\ion{N}{ii}]\,$\lambda$6583 as kinematical tracer since its 
higher mass minimizes thermal broadening. 
Only three target galaxies  with  nitrogen emission of $\rm{> 2 \times 
10^{-16}\,erg\ s^{-1}\,cm^{-2}}$\,\AA$^{-1}$ have been used  for this plot.
We point out that we did not correct for effects of kinematical line 
broadening. Thus, the data in Fig.~17 do not show the pure thermal line 
broadening. In order to establish nevertheless a relationship between 
ionization  and gas dynamics we minimized the effect of kinematical line 
broadening in using only slit positions located near the center of each galaxy
 (s1 in NGC\,1963, s1 in NGC\,3044 and s2 in NGC\,4634, cf. Fig.~\ref{fig1}). 
A correction for the instrumental profile has been applied.

\begin{figure}[t]
\begin{center}
\psfig{file=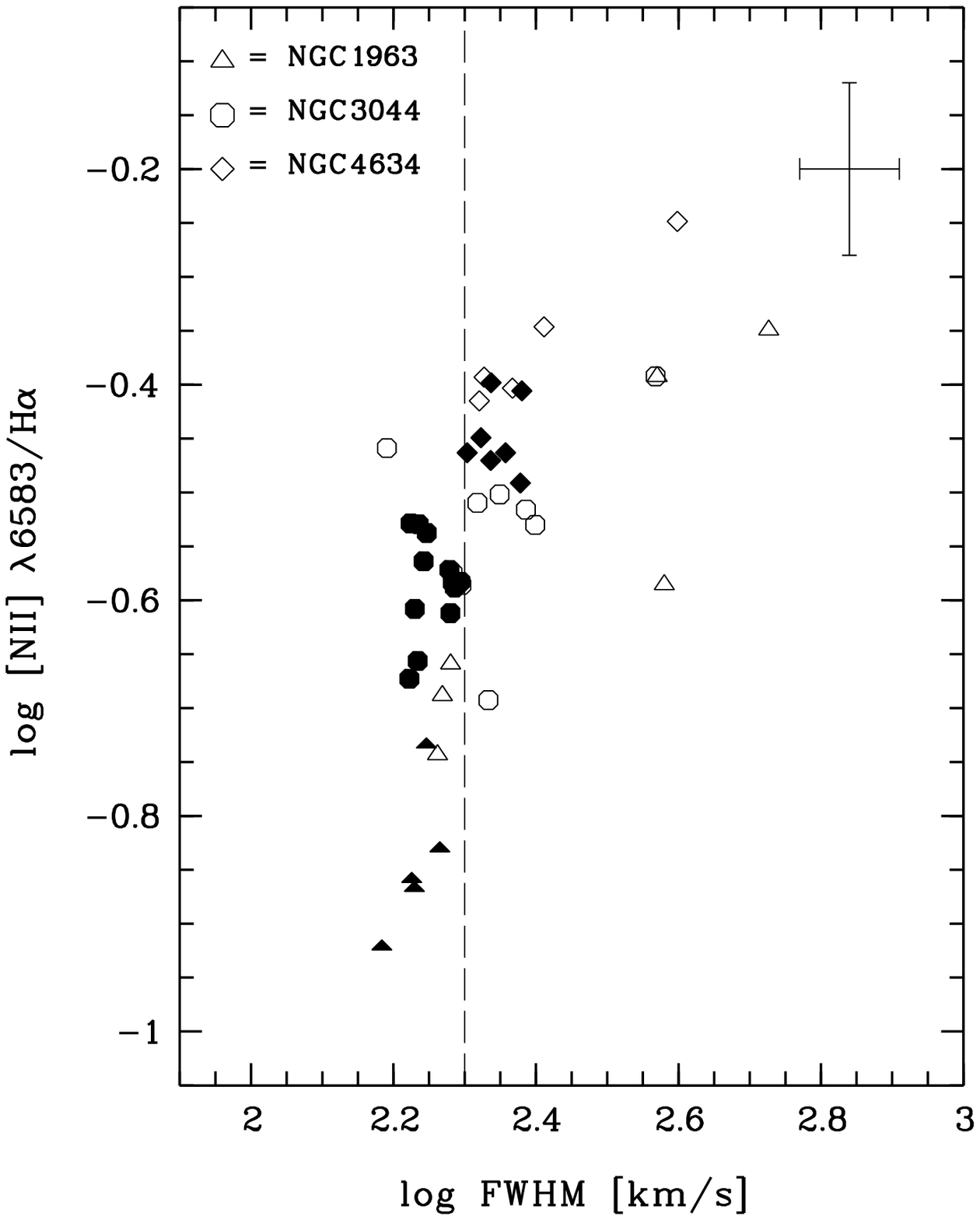,height=7.6cm,width=7.6cm,angle=0}
\caption{Relationship between [\ion{N}{ii}]\,$\lambda$6583/$\rm{H\alpha}$ 
and the corresponding [\ion{N}{ii}]-FWHM for all galaxies with 
broad nitrogen emission. Filled symbols correspond to the disk and open 
symbols denote the halo region. The cross represents the largest measured 
error for the halo area and the dashed line indicates the value of the 
kinematical resolution.}
\end{center}
\label{fig15}
\end{figure}
The spectral resolution has been determined to be 4.6\AA\ by measuring the 
FWHM at different positions of HeAr calibration spectra. This translates into 
210 km\,$\rm{s^{-1}}$ for the $\rm{H\alpha}$ emission line, corresponding to 
2.3 on a logarithmic scale. Because most data in Fig.~17 are located between 
2.2 and 2.4 one has to be aware that only values larger than 2.3 (within the 
errors) have a significant physical meaning. 
We therefore checked for consistency in measuring FWHM of the 
[\ion{O}{i}]$\lambda$6300 nightsky line in NGC\,1963 and NGC\,3044. Since 
this prominent emission feature is nearest to the 
[\ion{N}{ii}]\,$\lambda$6583 spectral line one thus obtains more precise 
values of the kinematical resolution. In the present case the parameter has 
to be modified slightly to 4.2\,\AA\ or 200\,km\,$\rm{s^{-1}}$, leading again 
to the same value as mentioned above.   
Fortunately we are interested in line broadening towards the halo and hence 
the main body of the data can be used for analysis.

For all galaxies plotted in Fig.~17 a clear correlation between gas kinematics
 and its ionization is visible. The broadening of lines at high galactic 
latitudes, even if less significant, supports a prominent large scale motion 
or increased turbulence of the ionized gas particles. 
Since this motion is tightly correlated to the ionization of the 
DIG it seems that in the halo of these galaxies shocks, superwinds or 
turbulent mixing layers as additional ionization or heating sources occur.

This can be seen very nicely in NGC\,1963 where disk and halo 
data cover different regions.
Additionally, the hybrid character of NGC\,3044 becomes obvious, again. 
Line broadening suggests only very moderate shock contributions. 
Although for NGC\,4634 the data allows no detailed diagnostics concerning 
possible ionization mechanisms Fig.~17 suggests that shocks (besides 
photoionization) could represent one possible additional excitation 
mechanism of the extraplanar DIG.

\section{Summary and conclusions}
Line ratios for the halos of NGC\,1963, NGC\,3044 and NGC\,4634 can be 
explained as a combination of photoionization by O stars and shock ionization. 
Since the measured FWHM increases with height above the galactic plane the 
results obtained from Fig.~17 support our findings that shocks can contribute 
to the ionization of the extraplanar DIG.
The photoionization models from Mathis and Dom\-g\"or\-gen\,\&\,Mathis are not
 always capable to reproduce all line ratios correctly (high  
[\ion{N}{ii}]\,$\lambda$6583/$\rm{H\alpha}$, 
[\ion{O}{i}]\,$\lambda$6300/$\rm{H\alpha}$, and  
[\ion{He}{i}]\,$\lambda$5876/$\rm{H\alpha}$). These models were 
developed to describe the measured line ratios for the Milky Way 
(Reynolds \cite{rey85a},b). However the local environment near 
the sun is not necessarily representative for the more active regions in our 
target galaxies. Thus new models are required which should rest upon recently
published extragalactic data and consider additional ionization/heating 
mechanisms such as shocks.

It is important to mention that the ratios for all galaxies in Fig.~16 are not 
constant over the whole range of $z$ and can vary significantly on small 
scales ($\approx$ 250\,pc). These results 
deviate from observations made e.g., by Golla et al. (\cite{gogode96}),
Rand (\cite{rand}), Otte\,\&\,Dettmar (\cite{otde}), or 
Haffner et al. (\cite{haff}).
Photoionization models can explain the general shape of 
$[\ion{S}{ii}]/[\ion{N}{ii}]$ but cannot account for individual changes 
(e.g., NGC\,4402).  
In addition non-linear changes, as visible in Fig.~16, correlate with density 
and are explainable by small scale density fluctuations.

If variations in $\rm{S^{+}}$ reflect variations in electron 
temperature, leading to $T_{e}= T_{e}(z)$, these fluctuations could be due to 
changing metal abundances or varying densities within the DIG. If different 
slopes in Fig.~15 or 
16 are indicators of different metallicities (Haffner et al. \cite{haff}), 
NGC\,4402 would have the same metallicity as the Milky Way or NGC\,891.
In accordance with Lehnert\,\&\,Heckman (\cite{leheck}) a correlation between 
line ratios (ionization) and line widths (gas kinematics) could be established.
Our results of a decrease of [\ion{O}{iii}]/H${\rm \beta}$ with increasing 
$|z|$ and a significant scatter in [\ion{S}{ii}]/[\ion{N}{ii}] are different 
compared to recent measurements in NGC\,891.   
All these aspects indicate that the ionization of the extraplanar DIG varies 
from galaxy to galaxy and within a galaxy. 

\begin{acknowledgements}
We thank D.J. Bomans, M. Rosa, H. Schulz, and R.A.M. Walterbos 
for helpful comments and enlightening discussions while writing the
paper. Many thanks also to H. Domg\"orgen for obtaining the data and special 
thanks to F. Valdes at NOAO for his valuable help during data reduction.
RT was supported by the DFG Graduiertenkolleg ``The Magellanic System and 
Other Dwarf Galaxies''. This paper was finalized while RJD enjoyed a 
sabbatical at ESO/Garching.
\end{acknowledgements}



\begin{thebibliography}{}
   \bibitem[1981]{bald} Baldwin J.A., Phillips M.M., Terlevich R., 1981,
      PASP 93, 5

   \bibitem[1984]{botti} Bottinelli L., Gouguenheim L., Paturel G.
	de Vaucouleurs G., 1984, A\&AS 56, 381

   \bibitem[1997]{dahl97} Dahlem M., Petr M.G., Lehnert M.D., Heckman T.M., 
     Ehle M., 1997, A\&A 320, 731    
   
   \bibitem[1992]{det92} Dettmar R.--J., 1992,
     Fund. Cosmic Phys. 15, 143
   
   \bibitem[1998]{det98} Dettmar R.--J., 1998, in:
      The Local Bubble and Beyond, Proceedings of the IAU Colloquium 
      No. 166, Breitschwerdt, D. Freyberg, M. J., Tr\"umper,
      J. (eds.), Springer LNP 506, p. 527

   \bibitem[1992]{deschu92} Dettmar R.--J., Schulz H., 1992,
     A\&A 254, 25 

   \bibitem[1991]{devau} de Vaucouleurs G., de Vaucouleurs A., Corwin J.R.,
	Buta R.J., Paturel G., Fouque P., 1991,
	Third reference catalogue of bright galaxies (1991)
	
   \bibitem[1997]{domade} Domg\"orgen H., Dettmar R.--J., 1997,
      A\&A 323, 391

   \bibitem[1994]{doma} Domg\"orgen H., Mathis J.S., 1994,
      ApJ 428, 647

   \bibitem[1996]{gogode96} Golla G., Dettmar R.--J., Domg\"orgen H., 1996,
      A\&A 313, 439  

   \bibitem[1997]{green97} Greenawalt B., Walterbos R.A.M., Braun R., 1997,
      ApJ 483, 666  

   \bibitem[1999]{haff} Haffner L.M., Reynolds R.J., Tufte S.L., 1999,
      ApJ 523, 223

   \bibitem[1989]{laube} Lauberts A., Valentijn E.A., 1989,
	The Surface Photmetry Catalogue of the ESO-Uppsala Galaxies

   \bibitem[1995]{lehe} Lehnert M.D., Heckman T.M., 1995,
    ApJS 97, 89

   \bibitem[1996]{leheck} Lehnert M.D., Heckman T.M., 1996,
    ApJ 462, 651

   \bibitem[1997]{martin} Martin, C.L., 1997, ApJ 491, 561

   \bibitem[1986]{ma} Mathis J.S., 1986,
    ApJ 301, 423
   
   \bibitem[1993]{mico} Miller W.W., Cox D.P., 1993,
    ApJ 417, 579
    
   \bibitem[1989]{oster} Osterbrock D.E., 1989,
      Astrophysics of Gaseous Nebulae and Active Galactic Nuclei,
      University Science Books

   \bibitem[1999]{otde} Otte B., Dettmar R.-J., 1999,
      A\&A 343, 705
       
   \bibitem[1996]{rand96} Rand R.J., 1996,
      ApJ 462, 712 
  
   \bibitem[1997]{rand97} Rand R.J., 1997,
      ApJ 474, 129
   
   \bibitem[1998]{rand} Rand R.J., 1998,
      ApJ 501, 137

   \bibitem[1985a]{rey85a} Reynolds R.J., 1985a,
      ApJ 294, 256

   \bibitem[1985b]{rey85b} Reynolds R.J., 1985b,
      ApJ 298, L30

   \bibitem[1993]{rey93} Reynolds R.J., 1993,
      AIP Conf. Proc. 278, Back to the Galaxy, ed. S.S. Holt \& F. Verter, 
   p.\,156
   
   \bibitem[1999]{rey99} Reynolds R.J., Haffner L.M., Tufte S.L., 1999,
      ApJ 525, L21
   
   \bibitem[2000]{rossa} Rossa J., Dettmar. R.-J., 2000, A\&A 359, 433
        
   \bibitem[1979]{shull} Shull J.M., McKee. C.F., 1979,
        ApJ 227, 131,

   \bibitem[1992]{teeri} Teerikorpi P., Bottinelli L., Gouguenheim L., 
	Paturel G., 1992, A\&A 260, 17

   \bibitem[1977]{tug} T\"ug H., 1977, ESO Messenger No. 11, p.7
   
   \bibitem[1987]{veio} Veilleux S., Osterbrock D.E., 1987,
     ApJS 63, 295

  
\end{thebibliography}
\end{document}